\documentclass[acmsmall,screen]{acmart}

\usepackage{multirow}
\usepackage{bm}
\usepackage{subfigure}
\usepackage{makecell}
\usepackage{tcolorbox}
\usepackage{soul}
\usepackage{amsmath}
\usepackage{ulem}
\usepackage{hyperref}
\usepackage{cleveref}
\usepackage{hyperxmp}

\newcommand{\phead}[1]{\vspace{1mm} \noindent {\bf #1}}
\newcommand\etal{{\it{et al.\ }}}

\newcommand{\yzz}[1]{\textcolor{black}{{#1}}}

\AtBeginDocument{%
  }

\setcopyright{acmlicensed}
\copyrightyear{2018}
\acmYear{2018}
\acmDOI{XXXXXXX.XXXXXXX}

\acmJournal{JACM}
\acmVolume{37}
\acmNumber{4}
\acmArticle{1}
\acmMonth{8}

\begin{document}

\title{An Empirical Study of Retrieval-Augmented Code Generation: Challenges and Opportunities}
% How do Retrieval Methods Enhance Pre-trained Models in Code Generation? An Empirical Study

\author{Zezhou Yang}
\email{yangzezhou@stu.hit.edu.cn}

\author{Sirong Chen}
\email{23S051017@stu.hit.edu.cn}

\author{Cuiyun Gao}
\authornote{Corresponding author.}
\affiliation{%
  \institution{Harbin Institute of Technology, Shenzhen}
%   \city{Shenzhen}
  \country{China}
  }
\email{gaocuiyun@hit.edu.cn}

\author{Zhenhao Li}
\email{zhenhao.li@ieee.org}
\affiliation{
  \institution{Concordia University}
  \city{Montreal}
  \country{Canada}
}

\author{Xing Hu}
\email{xinghu@zju.edu.cn}
\affiliation{
  \institution{Zhejiang University}
  \city{Hangzhou}
  \country{China}
}

\author{Kui Liu}
\email{brucekuiliu@gmail.com}
\affiliation{
  \institution{Huawei Technologies Co., Ltd.}
  \city{Hangzhou}
  \country{China}
}

\author{Xin Xia}
\email{xin.xia@acm.org}
\affiliation{
  \institution{Zhejiang University}
  \city{Hangzhou}
  \country{China}
}

\renewcommand{\shortauthors}{Z. Yang et al.}
% \renewcommand{\shorttitle}{How do Retrieval Methods Enhance Pre-trained Models in Code Generation Tasks}

\begin{abstract}
    Code generation aims to automatically generate code snippets of specific programming language according to natural language descriptions.
The continuous advancements in deep learning, particularly 
pre-trained models, have empowered the code generation task to achieve remarkable performance.
One main challenge of pre-trained models for code generation is the semantic gap between developers' natural language requirements and source code.
To address the issue, prior studies typically adopt a retrieval-augmented framework for the task, where the similar code snippets collected by a retrieval process can be leveraged to help understand the requirements and provide guidance for the generation process.
In a retrieval-augmented framework, similar data can be retrieved from the database using a retrieval algorithm, and original input data can be fused with retrieved data by different fusion strategies.
However, there is a lack of systematic study on the application of this framework for code generation, including the impact of the final generated results and the specific usage of the framework.
In this paper, we 
choose three popular pre-trained code models, namely CodeGen, UniXcoder, and CodeT5, to assess the impact of the quality and utilization of retrieved code on the retrieval-augmented framework.
Our analysis shows that the retrieval-augmented framework is beneficial for improving
the performance of the existing pre-trained models.
We also provide suggestions on the utilization
of the retrieval-augmented code generation framework: 
BM25 and Sequential Integration Fusion are recommended 
due to their convenience and superior performance. 
Sketch Filling Fusion, which extracts a sketch of relevant code, could help the model improve its performance further.
Additionally, we conduct experiments to investigate the influence of the retrieval-augmented framework on large language models for code generation, showing the effectiveness of the framework, and we discuss the trade-off between performance improvement and computational costs in each phase within the framework.
\end{abstract}




\maketitle

\section{Introduction}
\label{chap:chap1}
    %  Introduce background and code generation task
With the development of deep learning, pre-trained models have demonstrated remarkable performance in various code intelligence tasks.
These models are pre-trained on large-scale datasets including both code and text, and subsequently fine-tuned for specific downstream tasks~\cite{DBLP:conf/emnlp/FengGTDFGS0LJZ20,codexglue,codet5,unixcoder}.
The utilization of pre-trained models facilitates the effective resolution of a multitude of challenging tasks that are previously considered difficult.
Code generation task, which aims at automatically generating code based on natural language descriptions, is improved continuously by pre-trained models.
% Introduce code generation method
Prior researches predominantly employ Seq2Seq models to perform code generation tasks~\cite{DBLP:conf/acl/LingBGHKWS16,DBLP:conf/emnlp/IyerKCZ18}, some of which are augmented with structural information to bolster the syntactic correctness of the generated code~\cite{DBLP:conf/acl/RabinovichSK17, DBLP:conf/emnlp/IyerCZ19,DBLP:conf/emnlp/YinN18}.
These models can generate functional code on some simple datasets.
% The quality of the generated code cannot be guaranteed for more complex problems.
To handle the intricate development environments, researchers have introduced pre-trained models for code, which can achieve superior experimental results for code generation compared to previous models that are not pre-trained~\cite{codexglue,codet5,unixcoder}.
The task has proven effective in improving the efficiency of daily software development~\cite{intellicode,codex,pangu,codegeex}.  
Consequently, the task has attracted considerable attention among industry and academia, encouraging numerous practitioners and researchers to undertake a series of comprehensive studies~\cite{codex,alphacode,codegeex,nijkamp2023codegen,pangu,codet5+}.

% introduce the problems for current models and retrieved model
Recently, there have been studies proposing retrieval-augmented approaches to generate more accurate
% correct 
programs for the code generation task~\cite{DBLP:conf/nips/HashimotoGOL18,DBLP:conf/emnlp/ParvezACRC21,li2023skcoder,DBLP:conf/iclr/Zhou0XJN23}.
The relevant code snippets retrieved from a retrieval database are explicitly referenced as guideline within a code generation
model to enhance the generation performance
\cite{DBLP:conf/emnlp/ParvezACRC21}, and to improve the informativeness of the generated code~\cite{DBLP:journals/corr/abs-2104-05310}.
Despite the effectiveness, these retrieval-augmented approaches have not been adopted as a universal framework to help 
% multiple 
the existing models generate better code.
One main reason is that there is a lack of a comprehensive exploration on the utilization of the retrieval-augmented framework (RAF) for code generation.
% add an example
For example, noisy retrieved code snippets could decrease the model performance. 
In the RAF for code generation, relevant code snippets can be retrieved as reference by different retrieval techniques, and the original input can be augmented with the reference by various fusion strategies. Both the retrieval and fusion procedures can impact the model performance. It is necessary yet under-explored whether different code pre-trained models can benefit from the RAF for code generation. The impact of retrieval techniques and fusion strategies on the model performance are also worth investigating for achieving in-depth insights to researchers and practitioners.

To comprehensively explore the utilization of the RAF for code generation, we conduct extensive experiments in the paper.
Three popular pre-trained code models (i.e., CodeGen~\cite{nijkamp2023codegen}, UniXcoder~\cite{unixcoder} and CodeT5~\cite{codet5}) are evaluated on three widely-used
datasets (i.e., CONCODE~\cite{DBLP:conf/emnlp/IyerKCZ18}, CoNaLa~\cite{DBLP:conf/msr/YinDCVN08} and HearthStone~\cite{DBLP:conf/acl/LingBGHKWS16}).
Specifically,
we aim at answering the following three research questions (RQs):

{\noindent
\textbf{RQ1: What is the impact of retrieval-augmented framework on the performance of various  pre-trained models for code generation task?}
In this research question, we aim at investigating the impact of the RAF on the model performance and its generalization.
Without loss of generality, the basic text retrieval technique, BM25 ~\cite{BM25}, is adopted to retrieve relevant code snippets.
These retrieved code snippets are concatenated with
the original natural language description directly, serving as augmented data for the model. We fine-tune the pre-trained code models with the augmented data and demonstrate
that RAF is universally applicable to the
pre-trained models on 
different datasets for code generation. Experiment results show that the code generation performance of all the three models can benefit from the framework in various metrics. Especially on the HearthStone dataset, the three models exhibit an average improvement of
41.60\% in the EM metric, 9.01\% in the BLEU metric, and 8.69\% in the CodeBLEU metric.
}

{\noindent
\textbf{RQ2: How do the retrieval techniques
affect retrieval-augmented framework for code generation?}
In this research question, we explore how different kinds of retrieval techniques (i.e. code search models and text retrieval algorithms) in the RAF affect model performance.
The process that retrieves code snippets according to natural language can be regarded as code search task, so code search models can be used as retrieval techniques directly.
In this paper, we choose CodeBERT~\cite{DBLP:conf/emnlp/FengGTDFGS0LJZ20} and CoCoSoDa~\cite{shi2022cocosoda} as code search models to retrieve similar code snippets from training set based on the input natural language descriptions.
Popular text retrieval techniques such as BM25~\cite{BM25} and RetroMAE~\cite{DBLP:conf/emnlp/XiaoLSC22} are also involved.
Our experiment results show that all three models achieved the highest improvement from the retrieved results of BM25 on CONCODE
and HearthStone.
On CoNaLa, BM25 shows
optimal performance for CodeT5, and is suboptimal for CodeGen, which further demonstrates that the effectiveness of BM25 for the RAF.}

{\noindent
\textbf{RQ3: What is the impact of different strategies for fusing the retrieved results on the model performance?}
% How to better utilize the retrieved results for the generation process?}
In this research question, we study how to better integrate the retrieved code snippets for the code generation task from two aspects, including the number of retrieved results and the fusion strategies. For the fusion strategies, besides sequentially integrating the retrieved results with the natural language description, we also consider the Sample Expansion Fusion, Vectorized Decoding Fusion~\cite{DBLP:conf/eacl/IzacardG21}, and Sketch Filling Fusion~\cite{li2023skcoder}.
Our experiment results show that the quantity of retrieved code snippets should be determined according to the attributes of the specific dataset such as input/output length.
Sample Expansion Fusion with retrieved code snippets can improve the model performance to a great extent.
Based Sample Expansion Fusion, Sketch Filling Fusion yields an average improvement of 14.83\% in the BLEU metric and 8.05\% in the CodeBLEU metric across the three datasets for original CodeT5.
This improvement stands out as the highest among the four fusion strategies, suggesting that constructing sketches of relevant code could further enhance the model. 
However, the training associated with Sketch Filling Fusion is notably resource-intensive. In terms of balancing computational cost and performance enhancement, Sequential Integration Fusion proves to be a more cost-effective approach.
}

Through the large-scale empirical study, we achieve some findings and summarize the key findings as below.
\begin{enumerate}
    \item Retrieval-augmented framework could be adopted to improve the performance of various 
    % existing 
    pre-trained models for code generation.
    \item More complex retrieval techniques do not necessarily lead to better code generation results. BM25 is proven to be the most effective retrieval technique for code generation. 
    \item Sequential Integration Fusion with retrieved results is a simple but
    % and 
    effective fusion strategy. Despite the high computational costs, Sketch Filling Fusion can further improve the model performance.
\end{enumerate}

The major contributions of this paper are as follows:
\begin{enumerate}
    \item This paper serves as the first empirical study on the performance of retrieval-augmented framework for code generation. 
    \item We explore how different retrieval
    techniques and fusion strategies affect the performance of retrieval-augmented framework for code generation.
    \item We discuss the implications of our findings and provide actionable insights on
    % suggest further research on 
    the specific usage of the framework.
\end{enumerate}

The rest of the paper is organized as follows: Section \ref{chap:chap2} briefly introduces background and related work.
Section \ref{chap:chap3} elaborates the retrieval-augmented framework. 
Section \ref{chap:chap4} introduces the setup of our experiment study.
Section \ref{chap:chap5} presents the experiment results.
Section \ref{chap:chap6} discusses the implications of findings and threats to validity.
Section \ref{chap:chap7} presents the conclusions of the paper.
\section{Background and Related Work}
\label{chap:chap2}
    \subsection{Pre-trained Models in Code Intelligence}

Pre-trained models accumulate knowledge from large-scale unlabeled data through self-supervised training strategies, exhibiting superior generalization. Subsequent fine-tuning can yield commendable results on multiple tasks~\cite{gpt2}.
Encoder-only models are consistently used for code comprehension tasks with bidirectional attention.
CodeBERT\cite{DBLP:conf/emnlp/FengGTDFGS0LJZ20} is pre-trained on NL-PL pairs in six programming languages with Masked Language Modeling (MLM) and Replace Token Detection (RTD).
GraphCodeBERT~\cite{graphcodebert} designs pre-training tasks about data flow to improve code representation.
Decoder-only models are good at auto-regressive tasks like code generation and code completion with unidirectional attention.
CodeGPT~\cite{codexglue} is pre-trained on Java and Python corpus sourced from CodeSearchNet~\cite{codesearchnet}, which has the same architecture and training objective as GPT-2\cite{gpt2}.
Incoder~\cite{incoder}, employing a decoder-only architecture, demonstrates an impressive ability to predict tokens not only in a left-to-right sequence but also within the middle, utilizing information from both ends.
CodeGEN~\cite{nijkamp2023codegen} takes a conversational approach to program generation, where the process of code generation is described as multiple rounds of dialogue between the user and the system.
The step-by-step approach can break down long and complex intents into multiple simple intents, reducing the search space for models in each round of conversation.

In addition to employing either the encoder or the decoder of the Transformer~\cite{attention} independently, there are notable works that utilize the whole structure of Transformer in code intelligence, with UniXcoder~\cite{unixcoder} and CodeT5~\cite{codet5} standing out as popular models. 
UniXcoder, as a unified pre-trained model that incorporates semantic and syntax information from code comment and Abstract Syntax Tree (AST), can utilize mask attention matrices with prefix adapters to enable switching between Encoder-only, Decoder-only and Encoder-Decoder architecture.
CodeT5 is a unified pre-trained encoder-decoder model that leverages the token type information from code and has the ability to seamlessly support both code understanding and generation tasks.
These end-to-end architectures have been proved well-suited for various code-related tasks, including bug fixing, code translation, code summarization, and code generation~\cite{DBLP:journals/corr/abs-2302-04026}.

\subsection{Code Generation}
In recent years, code generation, a burgeoning research topic, has attracted widespread attention.
% traditional
Ling \etal ~\cite{DBLP:conf/acl/LingBGHKWS16} design a neural network architecture to convert the natural language description of a card into specific code implementation in Hearthstone or Magic the Gathering.
Yin \etal ~\cite{DBLP:conf/acl/YinN17} propose a novel approach powered by a grammar model which first translates natural language into an AST and subsequently regenerates it into code in a high-level programming language.
On this basis, many methods~\cite{DBLP:conf/acl/RabinovichSK17, DBLP:conf/emnlp/IyerCZ19, DBLP:conf/emnlp/YinN18, DBLP:conf/aaai/SunZXSMZ20} also leverage the AST of code to facilitate code generation in high-level programming languages.
Pre-trained models using code corpus have achieved excellent results in code generation, such as CodeGPT~\cite{codexglue}, PLBART~\cite{PLBART}, UniXcoder~\cite{unixcoder} and CodeT5~\cite{codet5}. The practice of pre-training followed by fine-tuning has become a mainstream pipeline to code generation.

% retrieval
For the code generation model, the decoding space of the model is large, which may lead to the poor quality of the generated code~\cite{codex}.
To alleviate such a problem, some work~\cite{DBLP:conf/emnlp/HayatiOAYTN18, DBLP:conf/emnlp/ParvezACRC21, DBLP:conf/nips/HashimotoGOL18, li2023skcoder, DBLP:conf/iclr/Zhou0XJN23} incorporates external knowledge to enhance the code generation model by means of retrieval. 
REDCODER~\cite{DBLP:conf/emnlp/ParvezACRC21} enhances code generation task by retrieving similar code snippets and enhances code summarization task by retrieving relevant text.
Hashimoto \etal ~\cite{DBLP:conf/nips/HashimotoGOL18} design a retrieve-and-edit framework, which retrieves the code first and then edits the retrieved results as the output code.
SKCODER~\cite{li2023skcoder} extracts templates from retrieved results as sketch, and edits the sketch into the output code.
DocPrompting~\cite{DBLP:conf/iclr/Zhou0XJN23} maintains a document library that stores the specification documents of the code, and enhances the code generation model by retrieving the document library. 
In addition to enhancing generation through retrieval, some work~\cite{chen2023codet, CompilerFeedback, le2022coderl} reconstructs the input and output of the model via pre-processing and post-processing techniques, resulting in higher-quality code. \yzz{Sepcically, CodeT \cite{chen2023codet}  allows for the selection of the optimal result within the candidate set of generated code by automatically generated test cases. CodeRL \cite{le2022coderl} determines the performance of the generated code through test cases, and uses this to supervise the training of the model. COMPCODER \cite{CompilerFeedback} incorporates both a generator and a discriminator to enhance the model's compilation capabilities.}

% -
% dataset
In the early works, semantic parsing datasets~\cite{ATIS, GEO} are commonly employed for evaluating the performance of code generation models, where HearthStone \cite{DBLP:conf/acl/LingBGHKWS16} is widely used due to its clear code implementation and structured requirement description.
Subsequently, the datasets specifically designed to mirror real-world programming scenarios come into prominence~\cite{DBLP:conf/acl/LingBGHKWS16, DBLP:conf/emnlp/IyerKCZ18, DBLP:conf/msr/YinDCVN08}. \yzz{For instance, CoNaLa, a dataset sourced from Stack Overflow, is meticulously constructed to serve as a benchmark for tasks that necessitate the generation of code from natural language descriptions.}
Currently, there is a wealth of diverse datasets in the field of code generation, with a primary emphasis on Python and Java. 
\yzz{CONCODE \cite{DBLP:conf/emnlp/IyerKCZ18} that is selected in CodeXGLUE \cite{codexglue} has been a comprehensive and necessary evaluation dataset for code generation.}
Furthermore, with the advancements in generative models, test case datasets~\cite{codex, APPS, MBPP, cassano2022scalable} have emerged, further assessing the code generation capabilities of models based on the pass@k metric~\cite{codex}.   
Simultaneously, there is a discernible shift from monolingual datasets towards multilingual ones~\cite{cassano2022scalable, zhu2022xlcost, wang2022mconala}, indicating a prevailing trend.

% -
% motivation
Due to the economic and technical potential of code generation, corporations continue to roll out pre-trained models for code generation~\cite{alphacode, pangu, codex}, and it can be found that the number of parameters and performance of these models continue to exceed conventional understanding.
When large language models (LLMs) demonstrate excellent code generation capabilities, code generation tasks have received more widespread attention and heated discussions. 
Several enterprises and institutions have launched a number of LLMs for code generation including CodeGeeX~\cite{codegeex}, AlphaCode~\cite{alphacode}, PanGu-Coder~\cite{pangu}, and CodeX~\cite{codex}.
With the continuous increase in the number of model parameters, the application ways are gradually diversified.
In particular, ChatGPT~\cite{ChatGPT} and GPT-4~\cite{gpt4} have strong generation capabilities in various fields, including code generation.
Many larger models of code are optimized by instruction-tuning in training stage~\cite{codet5+,codellama}.
In inference stage, in-context learning and chain-of-thought prompts are used to improve the generated results~\cite{selfplanCG,li2023enabling,li2023towards,bairi2023codeplan}.

\subsection{Retrieval-Augmented Generation}
Retrieval-augmented generation refers to improving the generation performance with the retrieved results provided by retrieval techniques.
For language models, the knowledge learned from training data is all stored in the parameters of the neural network.
The model might encounter challenges in generating the correct answer due to numerous parameters \cite{codex}.
Furthermore, when confronted with knowledge that has never been learned during pre-training, the model might fail to provide the accurate response.
The retrieved results can be regarded as a supplement to the implicit knowledge stored in the parameters of language models, encouraging the model to produce more accurate outputs~\cite{DBLP:conf/icml/GuuLTPC20}.
In addition, database can be modified and constantly updated, enabling the trained models to adapt to a broader range of new data~\cite{docprompt}.
In other words, retrieval-augmented generation achieves scalability of the modification or replacement of retrieval sources without the need to alter the models.

The k-Nearest Neighbor Language Model (kNN-LM)~\cite{knnLM} retrieve the k most similar training contexts for test context according to the distance in the embedding space of the pre-trained language model.
In fact, the k training contexts correspond to k training targets.
By normalizing and aggregating k training targets, kNN-LM can get a target distribution from k nearest neighbors, and the pre-trained language model can generate another target distribution directly according to current input.
kNN-LM can merge the two distributions above by weighted sum to get the final target distribution. 
Different from kNN-LM, Retrieval-Augmented Language Model (REALM)~\cite{DBLP:conf/icml/GuuLTPC20}, whose workflow can be summarized as the retriever-and-reader, has two components that are both trained. 
One is a neural knowledge retriever, which retrieves similar text with input using a dense inner product model.
The other is the knowledge-augmented encoder, which predicts the final results based on input and the retrieved text in the last step.
Actually, the prediction cannot generate texts using the encoder but extract a contiguous sequence from the retrieved text as the result.
A similar workflow has been proposed and developed~\cite{DBLP:conf/acl/ChenFWB17,DBLP:conf/naacl/NiZCM19} before REALM occurs.
To give more powerful ability to process different kinds of tasks, Lewis et al.
% Retrieval-augmented generation framework 
~\cite{DBLP:conf/nips/LewisPPPKGKLYR020} replaces reader with generator under the workflow of retriever-and-reader, which is also called retriever-and-generator.
Indeed, retrieval-augmented methods have been used in code generation and have developed for a long time.
Hashimoto \etal ~\cite{DBLP:conf/nips/HashimotoGOL18} proposes a retrieve-and-edit framework that edits the retrieved results to the desired output instead of generating code directly.
\yzz{The workflow of retrieve-and-edit is similar to retriever-and-generator~\cite{DBLP:conf/nips/LewisPPPKGKLYR020}, but the details of the two models are different.}
In specific, retrieval has more powerful components and can combine retriever with generator to fine-tune end-to-end.

Although the development of the models is rapid, fine-tuning following pre-training remains an indispensable paradigm for small-sized models.
Based on this mode, the utilization of similar retrieved results to improve pre-trained models for code generation has demonstrated effectiveness, attracting considerable attention.
However, most of these approaches only use single configuration, focusing on a specific view, and do not systematically summarize the retrieval-augmented framework and its usage from various aspects.
Therefore,
\yzz{we abstract three phases of the retrieval-augmented framework for code generation (retrieval phase, fusion phase and generation phase)}
and conduct a systematic study.
\yzz{For the retrieval phase, previous studies related to code search highlight that deep learning-based models can mitigate the semantic gap between query and code compared with statistics-based algorithms~\cite{codeserach_survey}. However, deep learning-based techniques rely on labeled data to learn the parameters of the models, which might bring additional training costs~\cite{DBLP:journals/tois/ZhaoLRW24}. These models may perform worse than statistics-based approaches under zero-shot settings~\cite{beir}.}
% 不同于代码搜索任务中代码检索库中仅有代码，代码生成的数据集往往以<自然语言，代码片段>对的形式出现，这就为代码检索提供了更多可能性.
\yzz{For the fusion phase, while many fusion strategies can be used in the retrieval-augmented framework from previous studies \cite{DBLP:conf/eacl/IzacardG21,li2023skcoder}, there is a lack of systematic exploration into the effectiveness and impact of the fusion strategies.}
\yzz{For the generation phase, our main concern is the impact of the retrieval-augmented framework on various pre-trained code generation models. }

\yzz{Recently, LLMs have shown impressive results in various downstream tasks across domains \cite{codex,codegeex,DBLP:conf/kbse/GaoWGWZL23}. However, they encounter challenges in real-world scenarios, such as hallucinations \cite{DBLP:journals/corr/abs-2311-05232}, outdated knowledge \cite{DBLP:journals/corr/abs-2312-10997}, and unclear reasoning processes \cite{DBLP:conf/emnlp/LabanKAFXJW23}. Retrieval-augmented generation can prompt LLMs to produce reliable and accurate results by integrating real-time factual knowledge retrieved from external databases. Considering its effectiveness for LLMs on other tasks, we additionally conduct experiments to explore its impact on LLMs for code generation, discussed in detail in Section \ref{chap:chap6}.}

\label{chap:3.2}
\begin{figure}[h]
  \centering
  \includegraphics[width=\linewidth]{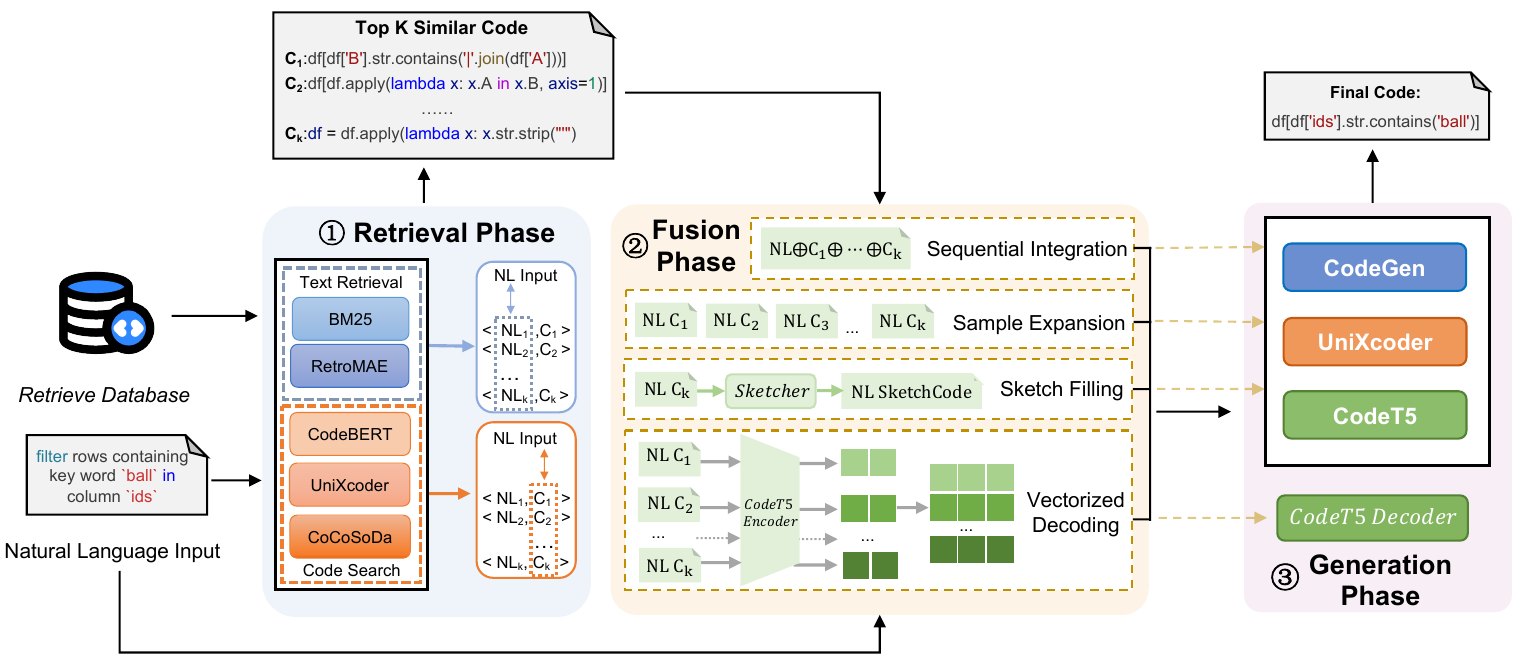}
  \caption{Overview of retrieval-augmented framework for code generation.}
  \label{fig:fig1-overview}
\end{figure}
\section{Retrieval-Augmented Framework}
\label{chap:chap3}
    \subsection{Overview}
\label{chap:3.1}

Figure~\ref{fig:fig1-overview} illustrates the overview of the general retrieval-augmented framework (RAF) for code generation. As can be seen, given a natural language input, the framework mainly includes three phases, i.e., \textit{Retrieval Phase}, \textit{Fusion Phase}, and \textit{Generation Phase}.
The details of each phase are as below. \textcircled{1}
\textit{Retrieval Phase}: The phase aims at retrieving
relevant code snippets based on the provided natural language description; \textcircled{2}
\textit{Fusion Phase}: In the phase, the retrieved code snippets are integrated
with original natural language description as augmented input for the generation phase.
\textcircled{3}
\textit{Generation Phase:} The phase leverages the augmented input to generate the code. All the phases can contribute to the model performance. Specifically, we first study the effectiveness and generalizability of the RAF by adopting various pre-trained code models in the \textit{Generation Phase}. We then explore the impact of different retrieval techniques in the \textit{Retrieval Phase} and fusion strategies in the \textit{Fusion Phase} on the model performance.

\subsection{Retrieval Phase}
In Retrieval Phase, the source of the retrieval database can be collected from GitHub, Stack Overflow, existing datasets and so on. 
The typical format of the retrieval database is <Natural Language Description, Code Snippet>.
There are two distinct ways to obtain the final retrieved code snippets.
The first way is regarding the database as a dictionary, where Natural Language Descriptions serve as the keys and corresponding Code Snippets act as the values. The text retrieval algorithms in Natural Language Processing (NLP) can measure the similarity between current natural language input and natural language description in database, and then lookup the corresponding code snippets as the final retrieved results.
From the other perspective, retrieving the relevant code snippets in database according to the natural language input meets the definition of code search. This means that various code search models can be used to retrieve relevant code snippets according to the natural language input.
In detail, code search models can directly measure the similarity between natural language input and code snippets in database, and return similar code snippets as retrieved results.
In this paper, we select five retrieval algorithms including two text retrieval algorithms (i.e. BM25~\cite{BM25} and RetroMAE~\cite{DBLP:conf/emnlp/XiaoLSC22}) and three code search models (i.e. CodeBERT~\cite{DBLP:conf/emnlp/FengGTDFGS0LJZ20}, UniXcoder~\cite{unixcoder} and CoCoSoDa~\cite{shi2022cocosoda}).

\subsubsection{Text Retrieval Algorithms}
There are two text retrieval algorithms in our experiments.
They retrieve similar natural language descriptions with natural language input, and select the corresponding code snippets as the retrieved results.

\textbf{BM25}~\cite{BM25} is commonly utilized for assessing the correlation between search terms and documents in document set.
Given a search term $Text$ and a document set $D$, $d \in D$ represents a document within the document set.
The document set $D$ serves as the search database, and within this database, the natural language in <Natural Language, Code Snippet> example pairs constitutes a document.
During the correlation calculation process, the search term $Text$, serving as natural language input, is initially divided into multiple tokens $\{t_{1}, t_{2}, \cdots, t_{i}, \cdots, t_{n} \}$.
The correlation score can be calculated as:  
\begin{equation}
    Score(T, d) = \sum^{n}_{i=1}(w_{i}R(t_{i}, d))
\end{equation}
where $w_{i}$ donates the weight of $t_{i}$, which is used to assess the importance of $t_{i}$ for document set $D$. 
$R(t_{i}, d)$ indicates the degree of correlation between $t_{i}$ and the document $d$.
$w_{i}$ is calculated as follows:
\begin{equation}
    w_{i} = IDF_{i} = log\frac{N-df_{i}+0.5}{df_{i}+0.5}
\end{equation}
where $N$ donates the number of documents in the database and $df_{i}$ donates the number of documents containing $t_{i}$.
$R(t_{i}, d)$ can be calculated according to the following formula:
\begin{equation}
    R(t_{i}, d) = \frac{f_{i}*(k+1)}{f_{i}+k*(1-b+b*\frac{L_{d}}{L_{avg}})}
\end{equation}
where $f_{i}$ donates the frequency value of $t_{i}$ in document $d$, $L_{d}$ donates the length of the document $d$, and $L_{avg}$ donates the average length of all documents. In the above formula, $b$ and $k$ are trainable parameters.

\textbf{RetroMAE}~\cite{DBLP:conf/emnlp/XiaoLSC22} is a pre-trained text retrieval model based on Masked Auto-Encoder (MAE) with asymmetric masking ratios for encoder and decoder.
It includes a full-scale BERT as encoder and a single-layer Transformer Decoder as decoder.
With the workflow of MAE, the input sentences are polluted with different masks and encoded as masked sentence embedding by encoder at first.
Subsequently, the decoder can recover the original sentence according to the masked sentence embedding.
Based on MAE, RetroMAE proposes an enhanced decoding process that can capture more training signals and diversified contexts with two-stream self-attention and position-specific attention mask. \yzz{In our experimental setup, we directly employ RetroMAE for inference because of the absence of labeled natural language data to train it for text retrieval tasks.}

\subsubsection{Code Search Models}
\label{sec:3.2.2}
There are three code search models in our experiments.
They can search the similar code snippets directly according to the natural language input. \yzz{To better adapt to the data distribution in the code generation domain, all three models are trained on the code search task using the training set of the corresponding code generation dataset.}

\textbf{CodeBERT}~\cite{DBLP:conf/emnlp/FengGTDFGS0LJZ20} is the first bimodal (i.e. natural language and programming language) pre-trained model for multiple programming languages with the same architecture as RoBERTa-base~\cite{DBLP:journals/corr/abs-1907-11692}.
For code search task, the natural language input and the code snippets are concatenated with "[CLS]" as the input of CodeBERT.
The representation of "[CLS]" is processed through a softmax layer to measure the semantic relevance between code and natural language input.

\textbf{UniXcoder}~\cite{unixcoder} is a unified pre-trained code model that is compatible with three different architectures (i.e. Encoder-Decoder model, Encoder-only model, and Decoder-only model), and controls the operation by mask attention matrices with prefix adapters.
To acquire semantic embedding of code, UniXcoder introduces two novel pre-training tasks: Multi-Modal contrastive Learning and Cross-Modal Generation.
Both of them can enhance the understanding ability of UniXcoder and improve the performance in downstream tasks.
For code search tasks, UniXcoder can utilize the encoder-only mode to separately encode natural language and code, followed by similarity computation.

\textbf{CoCoSoDa}~\cite{shi2022cocosoda} leverages contrastive learning for code search task and has achieved state-of-the-art performance up to now.
In detail, CoCoSoDa introduces four soft data augmentation and incorporates the momentum contrastive learning (MoCo) framework~\cite{DBLP:conf/cvpr/He0WXG20} to learn the representation of code and natural language query.
The core idea of contrastive learning is pulling similar representations and pushing apart discrepant representations.
Ultimately, both code and natural language query can be encoded into the high-dimensional vectors by code and query encoders designed using the same architecture as UniXcoder~\cite{unixcoder}, and their similarity can be measured with cosine similarity.

\subsection{Fusion Phase}
\label{chap:3.3}
Fusion Phase aims to connect the Retrieval phase and the Generation Phase.
If the code snippets retrieved by retrieval techniques are seen as the references that can provide additional knowledge, the fusion strategy in Fusion Phase is the guidance of the references for the subsequent code generation.
The fusion strategy utilizes the retrieved code snippets to alter the input while keeping the model architecture unchanged.
Here, we offer a brief introduction of four fusion strategies employed in Fusion Phase as follows:

\textbf{Sequential Integration Fusion (SIF):} Sequential Integration Fusion is a naive strategy to fuse retrieved code and natural language input. The k retrieved code snippets are seamlessly concatenated following the original natural language input by means of a special token <$retrieved\_code$> as new input. 

\textbf{Sample Expansion Fusion (SEF):} Sample Expansion Fusion refers to a strategy of enriching the original training dataset.
The k retrieved code snippets can be individually concatenated with their corresponding natural language input as k new instances, while maintaining the unchanged target code.
In other words, the training data can be expanded by creating new instances based on the retrieved results.
The volume of training data, except for the original data, is increased by a factor of k.
In the Generation Phase, the model can generate code according to both the original natural language input and the most similar code snippet.

\textbf{Vectorized Decoding Fusion (VDF):} 
% \todo{Add explaination}
Based on Sample Expansion Fusion that can transform k retrieved code snippets into k new instances, Vectorized Decoding Fusion encodes each new instance into a vector.
The k vectors \yzz{with the same natural language input} can be concatenated \yzz{along the hidden dimension} and utilized as input to the decoder, resulting in the final generation output.
In other words, Vectorized Decoding Fusion changes the fusion process by encoding code snippets into vectors, which can solve the problem of truncating and keep the information aggregated~\cite{DBLP:conf/eacl/IzacardG21}.
In our experiment, we use the encoder of CodeT5 to obtain the representation of new instances and fuse them in the decoder of CodeT5.

\textbf{Sketch Filling Fusion (SFF):} Vectorized Decoding Fusion can be considered as utilizing an encoder to extract the semantic information of the retrieved code snippets in a high-dimensional space.
However, Vectorized Decoding Fusion cannot explicitly capture similar structures in similar code.
Extracting the sketch of the most similar code snippet is advantageous for filtering out potentially irrelevant details and preserving the most useful and pertinent information.
The sketch can also be seen as a template, explicitly offering additional structure information to the model.
In specific, we use a neural encoder and a linear classification layer following SKCODER~\cite{li2023skcoder} to finish the sketch extraction of relevant code.

In addition to the specific fusion strategy, it should be considered how the number of code snippets used in the Fusion Phase impacts the model performance, which will be discussed in \ref{chap:5.3}.

\subsection{Generation Phase}
\label{chap:3.4}
% \todo{Abstractive Code Generation}
\yzz{In the Generation Phase, the generative model aims at generating the final code based on the data constructed during the Fusion Phase. The use of retrieved code snippets related to natural language input varies with the fusion strategy, while the architecture of the models remains unchanged. The original generative model presents the code generation tasks in the following format:} 
\begin{equation}
    F(<{x_1,x_2,...,x_n}>) = <{y_1,y_2,...,y_m}>
\end{equation}
\yzz{where $<{x_1,x_2,...,x_n}>$ is the natural language input and $<{y_1,y_2,...,y_m}>$ is the target code snippets. $F$ represents the process functionality of generative models, which can achieve the mapping from the natural language input into generated code snippets.}

\textbf{Code Generation with SIF.} \yzz{Sequential Integration Fusion is used to obtain an augmented dataset by appending similar retrieved code snippets after the original natural language input. It is necessary to introduce special separating tokens to help the generative model distinguish between the original input and the reference code snippets, thereby shifting the code generation process:}
\begin{equation}
    F(<{x_1,x_2,...,x_n}>,[retrieved\_code],C_1,[retrieved\_code],...,C_k) = <{y_1,y_2,...,y_m}>
\end{equation}
\yzz{where $[retrieved\_code]$ denotes the special separating tokens, and $C_k$ represents to the $k^{th}$ retrieved similar code snippets.}

\textbf{Code Generation with SEF.} \yzz{Contrasting with Sequential Integration Fusion, Sample Expansion Fusion individually elaborates on each of the k retrieved code snippets after the original natural language input. Consequently, it is employed in the following manner for k retrieved code snippets during the training stage:}
\begin{equation}
\begin{aligned}
\begin{cases}
    F(<{x_1,x_2,...,x_n}>,[retrieved\_code],C_1) & =  <{y_1,y_2,...,y_m}> \\
    & \hdots \\
    F(<{x_1,x_2,...,x_n}>,[retrieved\_code],C_k) & = <{y_1,y_2,...,y_m}>
\end{cases}
\end{aligned}
\end{equation}
\yzz{During the inference stage, the model generation process is subsequently represented as:}
\begin{equation}
    <{\hat{y_1},\hat{y_2},...,\hat{y_m}}> = F(<{x_1,x_2,...,x_n}>,[retrieved\_code],C) \\
\end{equation}
\yzz{where $C$ denotes the most similar code snippets retrieved from the codebase, and the models can generate the code prediction $<{\hat{y_1},\hat{y_2},...,\hat{y_m}}>$ based on the natural language input and $C$.}

\textbf{Code Generation with VDF.} \yzz{Vectorized Decoding Fusion involves encoding each instance from the Sample Expansion Fusion into a vector. These vectors with the same natural language input are concatenated and then fed into the decoder. This fusion process necessitates both an encoder and a decoder:}
\begin{equation}
\begin{aligned}
    \begin{cases}
    v_{C_1}& = Encoder(<{x_1,x_2,...,x_n}>,[retrieved\_code],C_1) \\
     & \hdots \\
    v_{C_k}& = Encoder(<{x_1,x_2,...,x_n}>,[retrieved\_code],C_k) \\
    \end{cases}\\
\end{aligned}
\end{equation}

\begin{equation}
    Vector_{Input} = concat([v_{C_1},...v_{C_k}]) 
\end{equation}

\begin{equation}
    Decoder(Vector_{Input}) = <{y_1,y_2,...,y_m}>  
\end{equation}

\textbf{Code Generation with SFF.} \yzz{Sketch Filling Fusion extracts the sketch of the retrieved similar code snippets, necessitating that the model edits the sketch to obtain the final code:}
\begin{equation}
    \begin{cases}
    F(<{x_1,x_2,...,x_n}>,[code\_sketch],S_1) & = <{y_1,y_2,...,y_m}> \\
    & \hdots \\
    F(<{x_1,x_2,...,x_n}>,[code\_sketch],S_k) & = <{y_1,y_2,...,y_m}>
    \end{cases}
\end{equation}
\yzz{where $[code\_sketch]$ donotes the special separating tokens, and $S_k$ represents to the sketch of $k^{th}$ retrieved similar code snippets. The inference stage is similar to Sample Expansion Fusion:}
\begin{equation}
    <{\hat{y_1},\hat{y_2},...,\hat{y_m}}> = F(<{x_1,x_2,...,x_n}>,[retrieved\_code],S) \\
\end{equation}
\yzz{where $S$ denotes the sketch of the most similar code snippets retrieved from the codebase.}

\section{Experiment Study Setup}
\label{chap:chap4}
    
\subsection{Dataset}
\label{chap:4.1}
We evaluate the RAF for code generation on three well recognized datasets: CONCODE~\cite{DBLP:conf/emnlp/IyerKCZ18}, CoNaLa~\cite{DBLP:conf/msr/YinDCVN08} and HearthStone~\cite{DBLP:conf/acl/LingBGHKWS16}.
The statistics of these three datasets are summarized in Table \ref{tab:dataset}.

{\parindent0pt
\textbf{CONCODE~\cite{DBLP:conf/emnlp/IyerKCZ18}} in CodeXGLUE~\cite{codexglue} is one of the most popular datasets for code generation task.
It includes about 33,000 Java projects collected on GitHub.
According to the GitHub repository, CONCODE can be divided into 100,000 instances for training and 4,000 instances for validation and testing.
The repository-based partitioning keeps the domain in the test set separate from the training set, which helps test the real generalization of models for unseen natural language descriptions. 
Each instance is a tuple of natural language descriptions, code environments, and code snippets, where the code environment includes other member variables and member functions in the class.
% An example in CONCODE is shown in Figure \ref{fig:fig2-concode}.
}

\begin{table*}
  \caption{Statistics of the datasets for code generation.}
  \label{tab:dataset}
  \begin{tabular}{cccccccc}
    \toprule
    \textbf{Dataset} & \textbf{Train/Test/Validation} & \makecell[c]{\textbf{Programming} \\ \textbf{Language}} & \makecell{\textbf{Max-Avg-Min length of} \\ \textbf{Input/Output}}\\
    \midrule
    CONCODE & 100,000/2,000/2,000 & Java & 2,246/264 - 213/33 - 18/6 \\
    CoNaLa & 2,179/500/200 & Python & 62/84 - 16/16 - 1/1 \\
    HearthStone & 533/66/66 & Python & 115/636 - 74/131 - 54/78 \\
  \bottomrule
\end{tabular}
\end{table*}

{\parindent0pt
\textbf{CoNaLa}~\cite{DBLP:conf/msr/YinDCVN08} comprises 2,879 manually annotated questions along with their corresponding Python solution instances from Stack Overflow. These instances encompass genuine natural language queries posed by programmers with diverse intentions. The length of input and output are shorter compared to the other two datasets as shown in Table \ref{tab:dataset}.
% An example of CoNaLa is shown in Figure \ref{fig:fig3-conala}.
}

{\parindent0pt
\textbf{HearthStone}~\cite{DBLP:conf/acl/LingBGHKWS16} is a collection of Python classes implemented for the HearthStone card game, containing 665 different HearthStone cards.
Each card contains a set of fields delineating the card information and Python code snippets implementing its corresponding functions.
These fields include semi-structured descriptions such as the card name, cost, attack, description, and other attributes.
Since most of the fields are similar among the cards, the code structures of different cards are comparable.
% The length of output is the largest in the three datasets.
% An example in HearthStone is shown in Figure \ref{fig:fig4-hs}.
}

% In summary, the three datasets represent diverse development environments for code generation.
We split the CONCODE and HearthStone datasets into training, test and validation sets by following the original papers.
Although CoNaLa has already been split into 2,379 training and 500 test instances, there are no validation instances available for experiments.
To facilitate our experimentation, we select 200 instances randomly from training data as the validation set for CoNaLa.

\subsection{Pre-trained Code Models}
\label{chap:4.2}
In terms of the model architecture, pre-trained models in code intelligence can be categorized into three types: Encoder-Decoder, Encoder-only and Decoder-only.
The models with the architecture of Encoder-only cannot finish generation tasks due to its inherent bidirectional representation.
In our experiments, three existing pre-trained code models are chosen to verify the effectiveness of the RAF.
These three models are CodeGen~\cite{nijkamp2023codegen}, UniXcoder~\cite{unixcoder}, and CodeT5~\cite{codet5}.
In the three models, CodeGen is a Decoder-only model, and the other two models (i.e. UniXcoder and CodeT5) have both Encoder-Decoder architecture.
They stand for different training strategies and generation processes, and the details are introduced as below.
Table~\ref{tab:baseline} presents the overview of these three pre-trained code models.

\textbf{CodeGen~\cite{nijkamp2023codegen}}, a kind of auto-regressive transformer, has a similar architecture to GPT-NEO~\cite{gptneo}.
The training objective is to maximize the likelihood of the target sequence given the context with a natural language corpus and programming language data collected from GitHub.
CodeGen comprises three versions according to sequential training datasets.
In short, CodeGen-NL is trained on The Pile.
CodeGen-MULTI continues training on BigPython based on CodeGen-NL, and CodeGen-MONO continues training on BigPython based on CodeGen-MULTI.
In our experiment, we choose the last version (i.e. CodeGen-MONO) due to its superior performance.

\begin{table}
  \caption{Overview of the pre-trained code models in the generation phase.}
  \label{tab:baseline}
  \begin{tabular}{cccccc}
    \toprule
    \textbf{Model} & \textbf{Parameters} & \textbf{Pre-trained Data} & \textbf{Input Length} & \textbf{Output Length}\\
    \midrule
    CodeGen & 350M & The Pile, BigQuery, BigPython & 2,048 & - \\
    UniXcoder & 126M & CodeSearchNet & 350 & 150\\
    CodeT5 & 223M & CodeSearchNet & 512 & 512\\
  \bottomrule
\end{tabular}
\end{table}

\textbf{UniXcoder~\cite{unixcoder}} can utilize the encoder-decoder mode for code generation tasks. Other detailed information of UniXcoder has been described in Section \ref{sec:3.2.2}.

% {\parindent0pt
\textbf{CodeT5~\cite{codet5}} 
based on T5~\cite{DBLP:journals/jmlr/RaffelSRLNMZLL20}, is a pre-trained model that can accomplish various code intelligence tasks through generation forms, much like how T5 is used for NLP tasks.
CodeT5 fully considers code-specific sequence and structural information with Identifier-Aware Denoising Pre-training in pre-training stage.
In the subsequent stage of pre-training, CodeT5 leverages NL-PL bimodal data for dual generation to close the gap between discrete knowledge from pre-training and continual knowledge from fine-tuning.
In a recent empirical study~\cite{DBLP:journals/corr/abs-2302-04026}, CodeT5 has been demonstrated as one of the most potent pre-trained models for code generation task.
% }

\subsection{Metrics}
\label{4.2}
% {\parindent0pt
\textbf{Exact Match Accuracy} (EM) represents the percentage of exact matches between predicted code and reference code (i.e. ground-truth), which shows that this metric is the most restrictive.
The metric is to measure the ability of the model to generate identical code, defined as:
% }
\begin{equation}
  EM = \frac{\sum^{|D|}_{i=1}(y_{i}==\hat{y_i})}{|D|}
\end{equation}

% {\parindent0pt
\textbf{BLEU}~\cite{DBLP:conf/acl/PapineniRWZ02} is an important metric to evaluate the quality of machine translation, and it is also widely used in other generation tasks. 
BLEU compares the n-gram in the generated code to measure the similarity with the reference code, where n-gram refers to the consecutive n tokens in the sentence.
In the code generation task, n is typically taken as 4, and we take BLEU-4 as one of our metrics to evaluate the RAF.
To ensure the fairness of evaluation, the calculation of BLEU metric needs to introduce a penalty term for n-gram operation. BLEU can then be expressed as the product of n-gram weighting and the penalty term. 
The calculation process is shown as follows:
% }
\begin{equation}
    BLEU = BP \cdot exp\left(\sum^{N}_{n=1}\omega_{n} \log{p_{n}}\right)
\end{equation}
\begin{equation}
    BP=
    \begin{cases}
        1 & \text{if c>r} \\
        e^{1-\frac{r}{c}} & \text{if c<=r}
    \end{cases}
\end{equation}
where $p_{n}$ means the modified n-gram precision and $\omega_{n}$ is the weight. $BP$ represents the brevity penalty. $c$ is the length of the output code, and $r$ is the length of the reference code.

\yzz{\textbf{Edit Distance} (ED) measures the syntactic similarity by the minimum number of single token edits required to transform predicted code snippets to target code snippets. The allowed edit operations include insertion, deletion, and substitution of tokens. Define \( ED(i, j) \) as the edit distance between the first \( i \) tokens of \( y \) and the first \( j \) tokens of \( \hat{y} \). The edit distance \( ED(i, j) \) can be computed using the following recurrence relation:}

\begin{equation} 
% \color{blue}
ED(i, j) = 
\begin{cases} 
\hspace{0.5cm} 0 &  \text{if } i = 0 \text{ and } j = 0 \\
\hspace{0.5cm} i &  \text{if } j = 0 \\
\hspace{0.5cm} j &  \text{if } i = 0 \\
\min 
\begin{cases} 
ED(i-1, j) + 1 \\
ED(i, j-1) + 1 \\
ED(i-1, j-1) + \delta(y[i-1], \hat{y}[j-1])
\end{cases} & \text{otherwise}
\end{cases} 
\end{equation}
\yzz{where $\delta(a, b)$ is defined as:}
\begin{equation} 
% \color{blue}
\delta(a, b) = \begin{cases} 
0 & \text{if } a = b \\
1 & \text{if } a \neq b.
\end{cases} 
\end{equation}

\yzz{\textbf{Similarity}}$\yzz{\bf{_{AST}}}$ \yzz{refers to the syntactic abstract syntax tree (AST) matching score to evaluate the structural information between predicted code snippets and the target code snippets. This is the formula to compute AST similarity:}
\begin{equation} 
% \color{blue}
    Similarity_{AST} = 1 - \frac{\text{Tree Edit Distance}(T, \hat{T})}{\max(Size(T), Size(\hat{T}))}
\end{equation}
\yzz{where $T$ and $\hat{T}$ represent the AST of predicted code snippets and the target code snippets separately. The computation of tree edit distance bears similarity to that of edit distance, with the primary distinction being that tree edit distance is calculated based on tree nodes rather than individual tokens.}

% {\parindent0pt
\textbf{CodeBLEU}~\cite{codebleu} is an improved metric based on BLEU by considering syntax and semantic information about code.
In specific, CodeBLEU incorporates the advantages of n-gram matching in BLEU, and is enhanced by code syntax with AST and code semantics with data flow:
\begin{equation}
    CodeBLEU = \alpha \cdot BLEU +\beta \cdot BLEU_{weight} +\gamma \cdot Similarity_{AST} + \delta \cdot Similarity_{DF}
\end{equation}
where $\alpha, \beta, \gamma, \delta$ are weight coefficients to control the percent of different metrics.
In contrast to assigning the same weight to each token in the BLEU calculation, $BLEU_{weight}$ assigns distinct weights to different tokens to obtain the n-gram matching score.
$Similarity_{AST}$ is syntactic AST similarity score, and $Similarity_{DF}$ is semantic data flow similarity score.

\subsection{Implementation Details}
\label{4.3}
All the pre-trained models and the corresponding tokenizer in our experiment are loaded from the official repository in Huggingface\footnote{https://huggingface.co/models}.
We adopt the PyTorch\footnote{https://pytorch.org/} framework to implement all the models and accomplish various tasks.
In our experiment, all the hyper-parameter settings of pre-trained models are the same as the original corresponding papers.
All the datasets are organized in the form of <Natural Language Description, Code Snippets> and stored in JSON files. Our computing devices are two Intel(R) Xeon(R) Platinum 8276 CPU @ 2.20GHz with 28 cores and two NVIDIA A100 (80G graphic memory in total).

The learning process of retrieval-augmented models aligns with the three phases illustrated in Figure \ref{fig:fig1-overview}.
To assess the effectiveness of the RAF without loss of generality, we conduct a series of experiments with BM25 and Sequential Integration Fusion in RQ1.
For RQ2, we leverage the five retrieval techniques outlined in Section \ref{chap:3.2} to retrieve the Top $k$ code snippets similar to the natural language input during Retrieval Phase. Subsequently, in Fusion Phase, we employ Sequential Integration Fusion to construct augmented data for different retrieved results.
In RQ3, we perform a controlled experiment to determine the optimal number of concatenated retrieved results, and then we employ the four different fusion strategies introduced in Section \ref{chap:3.3} in Fusion Phase.
\section{Empirical Study Results}
\label{chap:chap5}
\subsection{Research Questions}

We aim to answer the following research questions:
\begin{itemize}
\item[\textbf{RQ1:}] What is the impact of retrieval-augmented framework on the performance of various pre-trained models for code generation task?
\item[\textbf{RQ2:}] How do the retrieval techniques affect retrieval-augmented framework for code generation?
\item[\textbf{RQ3:}] What is the impact of different strategies for fusing the retrieved results on the model performance?
\end{itemize}

\begin{table*}
  \caption{Results comparison between the base model and the retrieval-augmented models. The base models are fine-tuned using the original datasets. The retrieval-augmented models are fine-tuned using the datasets augmented by Sequential Integration Fusion with the code snippets retrieved by BM25. Under each metric the best performance is marked as \textbf{bold}.}
  \label{tab:rq1}
  \resizebox{\linewidth}{!}{
  \begin{tabular}{c|ccccc|ccccc|ccccc}
    \toprule
     \multirow{2}*{\textbf{Model}} & \multicolumn{5}{c|}{\textbf{CONCODE}} & \multicolumn{5}{c|}{\textbf{CoNaLa}} & \multicolumn{5}{c}{\textbf{HearthStone}} \\
     ~ & EM & BLEU & \yzz{ED} & $\yzz{\text{Sim}_{\text{AST}}}$ & CodeBLEU & EM & BLEU & \yzz{ED} & $\yzz{\text{Sim}_{\text{AST}}}$ & CodeBLEU & EM & BLEU & \yzz{ED} & $\yzz{\text{Sim}_{\text{AST}}}$ & CodeBLEU \\
    \midrule
    CodeGen  & 19.55 & 21.83 & \yzz{19.13} & \yzz{36.80} & 27.33 & 9.00 & 12.74 & \yzz{9.48} & \yzz{22.15}  & 17.30 & 10.61  & 50.35 & \yzz{23.81} & \yzz{58.85} &  40.76 \\
    +BM25 & \textbf{21.20} & \textbf{24.99} & \yzz{\textbf{18.73}} & \yzz{\textbf{39.41}} & \textbf{30.04} & 8.60 & \textbf{15.58} & \yzz{\textbf{9.28}} & \yzz{\textbf{27.41}} & \textbf{22.24} & \textbf{15.15} & \textbf{54.35} & \yzz{\textbf{21.80}} & \yzz{\textbf{64.70}} & \textbf{44.00} \\
    UniXcoder  & 22.80 & 32.42 & \yzz{18.74} & \yzz{41.11} & 35.73 & 10.60 & 12.76 & \yzz{9.33} & \yzz{24.36} & 20.18 & 13.64 & 58.60 & \yzz{19.38} & \yzz{57.14} & 45.70 \\
    +BM25 & \textbf{23.50} & \textbf{35.80} & \textbf{\yzz{18.58}} & \textbf{\yzz{45.83}} &  \textbf{38.57} & \textbf{11.40} & \textbf{13.69} & \textbf{\yzz{9.05}} & \textbf{\yzz{25.46}} & \textbf{20.60} & \textbf{22.73} & \textbf{61.00} & \textbf{\yzz{16.90}} & \textbf{\yzz{61.92}} & \textbf{48.53} \\
    CodeT5 & 21.45 & 36.93 & \yzz{23.12} & \yzz{46.64} & 40.15 & 7.40 & 10.55  & \yzz{11.87} & \yzz{19.69} & 18.80 & 19.70 &  55.96 & \yzz{19.76} & \yzz{55.14} & 44.49 \\
    +BM25 & \textbf{23.35} & \textbf{40.42}  & \textbf{\yzz{21.91}} & \textbf{\yzz{47.67}} & \textbf{43.48} & \textbf{8.00} & \textbf{13.26} & \textbf{\yzz{10.46}} & \textbf{\yzz{23.29}} & \textbf{22.49} & \textbf{22.73} & \textbf{64.35} & \textbf{\yzz{14.95}} & \textbf{\yzz{63.65}} & \textbf{49.80} \\
    \bottomrule
  \end{tabular}
  }
\end{table*}

\subsection{RQ1: Effectiveness of Retrieval-Augmented Framework}

To broadly evaluate the way to leverage the retrieval-augmented framework into existing pre-trained models,
we study the effectiveness of the retrieval-augmented framework by comparing the performance of various pre-trained models before and after integration with the retrieved results on three distinct datasets: CONCODE~\cite{DBLP:conf/emnlp/IyerKCZ18}, CoNaLa~\cite{DBLP:conf/msr/YinDCVN08} and HearthStone~\cite{DBLP:conf/acl/LingBGHKWS16}. 

We present the comparison results in Table \ref{tab:rq1}.
% 整体阐述框架的有效性
Compared with fine-tuning using the original datasets, the retrieval-augmented framework achieves an average improvement of 6.79\%, 11.45\%, 6.93\%, and 8.72\% on CONCODE; 3.74\%, 18.42\%, 15.51\%, and 16.75\% on CoNaLa; 41.60\%, 9.01\%, 11.25\%, and 8.69\% on HearthStone for EM, BLEU, $\text{Similarity}_{\text{AST}}$ and CodeBLEU, respectively.
We perform a statistical significance test (t-test), and the results show that the models with the RAF outperform the original models at the significance level at 0.05 (p-value 0.035), demonstrating the effectiveness of the RAF on code generation.
% 整体阐述框架的泛化性
From the perspective of pre-trained models, within the RAF, the BLEU metric of CodeGen, UniXcoder, and CodeT5 are all increased by 14.90\%, 7.27\%, and 16.71\% on average across the three datasets, respectively. Above all, the experiment results indicate 
\yzz{that the RAF can improve the performance of various models within the same datasets, as well as identical models across different datasets,} showcasing its generalization.
In addition, the general improvements of multiple metrics, especially BLEU and CodeBLEU, can reflect the framework can help the model focus on the semantic and structural information of generated code during the code generation process.

% 根据数据集分析框架的效果
Specifically, on the HearthStone dataset, the retrieval-augmented framework yields notable enhancements in CodeGen by 42.79\%, UniXcoder by 66.64\%, and CodeT5 by 15.38\% in the EM metric. The other four metrics exhibit the most substantial improvements among the three datasets. The phenomena may be attributed to the regular code structure of HearthStone. With the similar field of cards, the code snippets have a similar form so that the relevant retrieved results can provide the field of a card to assist in implementing the code.
% 指出不同的模型都能受益于检索增强框架 综合来看CodeT5受到增益更大
Among three pre-trained code models, CodeT5 can improve 4.86 on the BLEU metric and 3.05 on the CodeBLEU metric across the three datasets. The highest comprehensive improvement suggests that CodeT5 can be enhanced to a greater extent comparing to the other models. As one of the most powerful pre-trained models for code generation~\cite{DBLP:journals/corr/abs-2302-04026}, CodeT5 still possesses the potential for further advancements in code generation within the RAF.

\begin{center}
\begin{tcolorbox}[colback=gray!10,%gray background
                  colframe=black,% black frame colour
                  width=\linewidth,
                  arc=1mm, auto outer arc,
                  boxrule=1pt,
                 ]
\textbf{Finding 1:} Retrieval-augmented framework is universal for various existing pre-trained code models to improve the code generation performance on different datasets effectively.
The utilization of retrieved code snippets in the framework can \yzz{assist models in focusing on the semantic and structural information of generated code.}

\end{tcolorbox}
\end{center}

\subsection{RQ2: Impact of Retrieval Techniques for Retrieval-Augmented Framework}
\label{chap:5.2}

\begin{table*}
  \caption{Results of the retrieval-augmented model on three datasets with different retrieval techniques. The percentages in parentheses following BLEU and CodeBLEU denote the enhancement achieved with various retrieval techniques compared to the baseline. Under each metric, the best performance is highlighted in \textbf{bold}.} 
  \resizebox{\linewidth}{!}{
  \begin{tabular}{c|c|cc|cc|cc}
    \toprule
    \multirow{2}{*}{\textbf{Model}} & \multirow{2}{*}{\makecell{\textbf{Retrieval}\\ \textbf{Technique}}} & \multicolumn{2}{c|}{\textbf{CONCODE}} & \multicolumn{2}{c|}{\textbf{CoNaLa}} &  \multicolumn{2}{c}{\textbf{HearthStone}} \\
    ~ & ~ &  \textbf{BLEU}  & \textbf{CodeBLEU} & \textbf{BLEU} & \textbf{CodeBLEU} &  \textbf{BLEU} & \textbf{CodeBLEU} \\
    \midrule
    \multirow{6}{*}{CodeGen} & baseline & 21.83 & 27.33 & 12.74 & 17.30 & 50.35 & 40.76 \\
    ~ & \textbf{BM25} & \textbf{24.99}($\bm{14.48\% \uparrow}$) & \textbf{30.04}($\bm{9.92\% \uparrow}$) & 15.58($22.29\% \uparrow$) & 22.24($28.55\% \uparrow$) & \textbf{54.35}($\bm{7.94\% \uparrow}$) & \textbf{44.00}($\bm{7.95\% \uparrow}$)\\
    ~ & RetroMAE & 20.14($7.74\% \downarrow$) & 26.18($4.20\% \downarrow$) & 12.93($1.49\% \uparrow$)& 21.17($22.37\% \uparrow$) & 9.40($81.33\% \downarrow$) & 16.93($58.46\% \downarrow$) \\
    ~ & CodeBERT & 24.18($10.77\% \uparrow$) & 29.23($6.95\% \uparrow$) & 13.61($6.83\% \uparrow$) & 23.23($34.28\% \uparrow$) & 52.08($3.44\% \uparrow$) & 42.07($3.21\% \uparrow$)\\
    ~ & \yzz{UniXcoder} & \yzz{24.76($13.42\% \uparrow$)} & \yzz{29.09($6.44\% \uparrow$)}& \yzz{14.63($14.84\% \uparrow$)} & \yzz{22.55($30.35 \% \uparrow$)} & \yzz{51.44($2.16\% \uparrow$)} & \yzz{43.59($6.94\% \uparrow$)}\\
    ~ & CoCoSoDa & 24.98($14.43\% \uparrow$) & 29.97($9.66\% \uparrow$) & \textbf{16.50}($\bm{29.51\% \uparrow}$) & \textbf{23.25}($\bm{34.39\% \uparrow}$) & 47.27($6.12\% \downarrow$) & 39.10($4.07\% \downarrow$) \\
    \midrule
    \multirow{6}{*}{UniXcoder} & baseline & 32.42 & 35.73 & 12.76 & 20.18 & 58.60 & 45.70 \\
    ~ & \textbf{BM25} & \textbf{35.80}($\bm{10.43\% \uparrow}$) & \textbf{38.57}($\bm{7.95\% \uparrow}$) & 13.69($7.29\% \uparrow$) & 20.60($2.08\% \uparrow$) & \textbf{61.00}($\bm{4.10\% \uparrow}$) & \textbf{48.53}($\bm{6.19\% \uparrow}$) \\
    ~ & RetroMAE & 32.02($1.23\% \downarrow$) & 35.47($0.73\% \downarrow$)& 13.77($7.92\% \uparrow$) & 20.65($2.33\% \uparrow$) & 55.89($4.62\% \downarrow$) & 44.45($2.74\% \downarrow$) \\
    ~ & CodeBERT & 35.15($8.42\% \uparrow$) & 38.13($6.72\% \uparrow$) & 13.61($6.67\% \uparrow$) & 20.73($2.73\% \uparrow$)& 56.12($4.23\% \downarrow$) & 45.28($0.92\% \downarrow$) \\
    ~ & \yzz{UniXcoder} & 34.79($7.31\% \uparrow$) & 37.71($5.54\% \uparrow$) & \yzz{13.32($4.39\% \uparrow$)} & \yzz{19.57($3.02\% \downarrow$)}& \yzz{55.79($4.80\% \downarrow$)} & \yzz{46.38($1.49\% \uparrow$)}\\
    ~ & CoCoSoDa & 34.05($5.03\% \uparrow$) & 37.38($4.62\% \uparrow$) & \textbf{14.55}($\bm{14.03\% \uparrow)}$)& \textbf{21.48}($\bm{6.44\% \uparrow}$) & 59.25($1.11\% \uparrow$) & 45.91($0.46\% \uparrow$)\\
    \midrule
    \multirow{6}{*}{CodeT5} & baseline & 36.93  & 40.15 & 10.55 & 18.80 & 55.96 & 44.49\\
    ~ & \textbf{BM25} & \textbf{40.42}($\bm{9.45\% \uparrow}$) & \textbf{43.48}($\bm{8.29\% \uparrow}$)& \textbf{13.26}($\bm{25.69\% \uparrow}$) & \textbf{22.49}($\bm{19.63\% \uparrow}$) & \textbf{64.35}($\bm{14.99\% \uparrow}$) & \textbf{49.80}($\bm{11.94\% \uparrow}$)\\
    ~ & RetroMAE & 37.90($2.63\% \uparrow$) & 42.30($5.35\% \uparrow$) & 11.28($6.92\% \uparrow$) & 21.76($15.74\% \uparrow$) & 61.11($9.20\% \uparrow$) & 48.32($8.61\% \uparrow$) \\
    ~ & CodeBERT & 39.00($5.61\% \uparrow$) & 43.05($7.22\% \uparrow$) & 12.53($18.77\% \uparrow$) & 21.18($12.66\% \uparrow$) & 62.09($10.95\% \uparrow$) & 49.09($10.34\% \uparrow$) \\
    ~ & \yzz{UniXcoder} & \yzz{38.21($3.47\% \uparrow$)} & \yzz{41.38($3.06\% \uparrow$)} & \yzz{12.73($20.66\% \uparrow$)} & \yzz{21.03($11.86\% \uparrow$)} & \yzz{54.23($3.09\% \downarrow$)} & \yzz{49.19($10.56\% \uparrow$)}\\
    ~ & CoCoSoDa & 38.93($5.42\% \uparrow$) & 42.72($6.40\% \uparrow$) & 12.24($16.02\% \uparrow$) & 21.77($15.80\% \uparrow$) & 62.51($11.70\% \uparrow$) & 48.87($9.84\% \uparrow$)\\
    \bottomrule
  \end{tabular}
  }
  \label{tab:tab4}
\end{table*}

In RQ1, our primary focus is to examine whether the RAF can improve the performance of various pre-trained code models.
According to the comparison of the results in Table \ref{tab:rq1}, the effectiveness and generalization of the RAF are validated.
In this research question, we assessed the ability of different retrieval techniques by comparing model performance before and after integration in the RAF for each of the three datasets separately, and 
the results are shown in Table \ref{tab:tab4}.

The results obtained from the retrieval-augmented model for code generation, employing various retrieval techniques, merit further investigation.
In the results of Table \ref{tab:tab4}, all three models achieve the highest generation gains from the retrieved results of BM25 on CONCODE and HearthStone.
Furthermore, CodeT5 achieves optimal performance when using the code snippets retrieved by BM25 on the CoNaLa dataset, with an improvement of 25.69\% in the BLEU score and a 19.63\% enhancement in the CodeBLEU score.
These two improvements are both the highest in three pre-trained code models on three datasets.
From this perspective, BM25, requiring no training, should be considered as the most promising 
retrieval technique to be explored within the RAF.
Apart from BM25, leveraging the retrieved results of CoCoSoDa, CodeGen and UniXcoder can enhance the performance of code generation maximally on CoNaLa,
which indicates the effectiveness of CoCoSoDa for certain models and datasets.

A noteworthy observation in Table \ref{tab:tab4} is that the performance of CodeGen and UniXcoder declined by 7.74\% and 1.32\% in the BLEU score after incorporating the retrieved results from RetroMAE on CONCODE.
Even though the performance of CodeT5 can improve to some extent, RetroMAE contributes the smallest improvement to model performance among the five retrieval techniques.
The results indicate that the code snippets retrieved by RetroMAE contribute to a marginal and even inverse improvement of model performance.
In other words, the retrieved results from RetroMAE are not always beneficial for the code generation process.
Coincidentally, Table \ref{tab:tab4} shows a similar scenario about RetroMAE on HearthStone.
As mentioned in Section \ref{chap:3.2}, RetroMAE is employed to retrieve code directly without fine-tuning for code generation \yzz{due to the absence of appropriate labeled natural language data}.
It is evident that RetroMAE performs subpar in this aspect on certain datasets, such as CONCODE and HearthStone, which feature distinct modes of expression compared to the field of NLP.
Conversely, RetroMAE demonstrates relatively better performance in CoNaLa, where the natural language input aligns more closely with the text found in the realm of NLP.
The substantial disparity between the code generation task and the text retrieval task could potentially explain why RetroMAE underperforms compared to the other four retrieval techniques.

According to the results in Table \ref{tab:tab4}, CodeBERT and UniXcoder are not typically suggested as retrieval techniques in Retrieval Phase for the RAF. As code search models, require fine-tuning on downstream datasets, which necessitates additional resources. While they outperform RetroMAE, they cannot attain a substantial improvement compared to BM25 and CoCoSoda. 

\begin{center}
\begin{tcolorbox}[colback=gray!10,%gray background
                  colframe=black,% black frame colour
                  width=\linewidth,
                  arc=1mm, auto outer arc,
                  boxrule=1pt,
                 ]
\textbf{Finding 2:} 
BM25 is suggested as a retrieval technique due to its consistently impressive performance
% improvements
and its inherent characteristic of requiring no training.
As the state-of-the-art code search model, CoCoSoDa could play an important role in enhancing existing pre-trained models. 

\end{tcolorbox}
\end{center}

\begin{figure}[h]
  \centering
  \includegraphics[width=\linewidth]{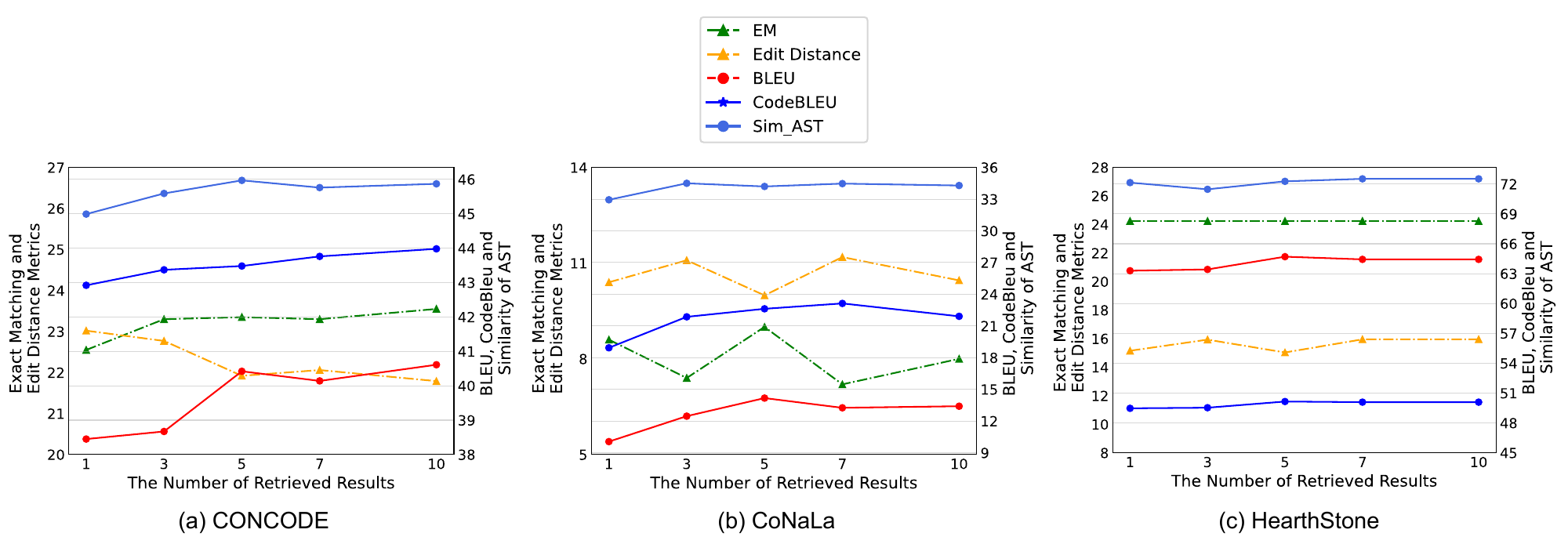}
  \caption{The impact of the number of retrieved code snippets using Sequential Integration Fusion on the effectiveness of CodeT5. The green line and yellow line represent the EM metric and Edit Distance metric corresponding to the left vertical axis. The red, blue and royalblue lines represent the BLEU metric, CodeBLEU metric and Simliarity of AST, respectively. All the three metrics correspond to the right vertical axis.}
  \Description{The impact of the number of retrieved code using Sequential Integration Fusion on the effectiveness of CodeT5.}
  \label{fig:fig5-number}
\end{figure}

\subsection{RQ3: The Usage of the Retrieved Results for the Code Generation}
\label{chap:5.3}

The effectiveness and generalization of retrieval-augmented models for code generation has been proved in RQ1, and the impact of different retrieval techniques has been discussed in RQ2.
In RQ3, we will further explore how to better utilize the retrieved results for code generation process.
Initially, our investigation focuses on assessing the impact of the number of retrieved results on the final performance of the retrieval-augmented models using Sequential Integration Fusion, as employed in RQ1 and RQ2.
Subsequently, several other fusion strategies are introduced to optimize the utilization of retrieved results, consequently enhancing the generation performance of the models.
In detail, owing to the wonderful and consistent performance of CodeT5 across three datasets in RQ2, we opt to use CodeT5 as the base model for this research question, which focuses on the utilization of retrieval results.

\subsubsection{The Impact of the Number of Utilized Retrieved Results}
\label{sec:5.4.1}
To explore the impact of the number of retrieved code snippets on the model improvement, we conduct experiments with Sequential Integration Fusion, as employed in RQ1 and RQ2. 
We employ CodeT5 to concatenate 1, 3, 5, 7 and 10 retrieved code snippets across three datasets and visualized the experimental results in a line chart as Figure \ref{fig:fig5-number} shown.

As the number of retrieved results increases on CONCODE, an overall upward trajectory is observed in the performance of retrieval-augmented CodeT5 across all three metrics.
At the same time, another important observation in Figure \ref{fig:fig5-number} (a) is the inflection point of the line for BLEU metric.
It occurs during the transition from concatenating three to concatenating five retrieval results.
On CoNaLa, a comparable trend is noticeable in Figure \ref{fig:fig5-number} (b), where the BLEU metric initially improves and then declines.
Similar patterns are observed in other metrics as well, suggesting that as the number of code snippets increases, there is an initial enrichment of input information followed by a gradual diminishment. 
In other words, larger number of retrieved results does not indicate a better improvement.
Considering that an increase in the length of input also leads to increased training costs, this inflection point could be a practical guideline for determining the optimal number of retrieved results.
The metrics on HearthStone present a distinct scenario compared to the other two datasets: all three metrics remain unchanged when five results are concatenated.
The reason is that the code snippets in HearthStone are all long, and over five code examples will be truncated by the model.
This can also be seen as one of the main drawbacks of the Sequential Integration Fusion.

\subsubsection{Comparison of Various Fusion Strategy}
In Section \ref{sec:5.4.1}, the limitation of Sequential Integration Fusion is raised on HearthStone.
In specific, the limitation is that only a finite number of retrieved results can be fused in the generation process of model, and the much longer augmented input will be truncated.
In this case, more fusion strategies should be explored to enrich the RAF.
According to the experiment results and findings in Section \ref{sec:5.4.1}, we choose 5 as the number of retrieved code in the next experiments. 

Table \ref{tab:fusion} shows the experiment results of the retrieval-augmented model for three datasets with four fusion strategies mentioned in Section \ref{chap:3.3}.
\yzz{Across all three datasets, Sequential Integration Fusion (SIF) and Sketch Filling Fusion (SFF) consistently deliver the best performance. On CoNaLa, SIF outperforms SFF due to the short input and output lengths, which do not constrain CodeT5's performance. Additionally, the data in CoNaLa lacks a consistent structure, rendering SFF less effective as it struggles to extract a beneficial code sketch for model generation.
Conversely, SFF excels on Hearthstone by capturing the analogous structure of similar code and effectively filtering out noisy variables.
Another noteworthy observation is that the performance of Sample Expainsion Fusion (SEF) on Hearthstone is surpassed only by SFF, and exceeds that of SIF. Moreover, the performance of SEF on the other two datasets closely rivals that of SFF. As outlined in Section \ref{chap:3.4}, the code generation process using SFF is akin to that of SEF, with the sole distinction being whether it is the code or the code sketch that is appended after the natural language input. Consequently, SEF can be employed to enhance the model's performance, particularly for small datasets such as Hearthstone, in a manner akin to data augmentation. Furthermore, SFF can build upon the improvements made by SEF to further enhance the model's performance.}
Vectorized Decoding Fusion (VDF) is also based on SEF. However, it does not bring additional enhancements to the model as SFF does. Specifically, VDF demonstrates decreased performance compared to SEF across three datasets.
The potential reason for this could be twofold: either the encoder of CodeT5 falls short in capturing a robust representation of the augmented input, or the decoder of CodeT5 struggles to generate code adequately based on multiple vector representation.

\begin{table}
  \caption{Results of the retrieval-augmented model for three datasets with various fusion strategies. "-" represents the base model fine-tuned on original datasets. Under each metric, the best performance is highlighted in \textbf{bold}. The training and inference costs represent the total training time and inference time for different fusion strategies separately.}
  \resizebox{\linewidth}{!}{
  \begin{tabular}{ccccccccc}
    \toprule
    \textbf{Dataset} & \textbf{Fusion Strategy} & \textbf{EM} & \textbf{BLEU} & \textbf{ED} &\textbf{$\text{Sim}_\text{AST}$} & \textbf{CodeBLEU} & \textbf{Training Costs} & \textbf{Inference Costs}\\
    \midrule
    \multirow{5}{*}{CONCODE} & - & 21.45 & 36.93 & 23.12 & 46.64  & 40.15 & 128 min & 547s \\
    ~ & Sequential Integration Fusion & \textbf{23.35} & 40.42 & 21.91 & \textbf{47.67} & \textbf{46.92} & 285 min & 763s\\ 
    ~ & Vectorized Decoding Fusion & 11.30 & 28.72 & 22.55 & 45.65 & 39.37 & 393 min & 1,662s \\
    ~ & Sample Expansion Fusion & 22.40 & 40.58 & 21.67 & 45.85 & 45.61 & 923 min & 643s \\
    ~ & Sketch Filling Fusion & 21.90 & \textbf{40.84} & \textbf{19.10} & 46.45 & 46.40 & 917 min & 805s \\
    \midrule
    \multirow{5}{*}{CoNaLa} & - & 7.40 & 10.55 & 11.87 & 19.69 & 17.25 & 13 min & 53s\\
    ~ & Sequential Integration Fusion & \textbf{8.00} & 13.26 & \textbf{10.46} & \textbf{23.29} & \textbf{21.46} & 18 min & 81s \\
    ~ & Vectorized Decoding Fusion & 7.20 & 10.91 & 13.49 & 21.50 & 17.69 & 49 min & 138s \\
    ~ & Sample Expansion Fusion & 1.80 & 15.91 & 13.92 & 10.78 & 17.95 & 47 min & 63s \\
    ~ & Sketch Filling Fusion & 1.80 & \textbf{17.01} & 13.55 & 12.36 & 18.65 & 46 min & 60s \\
    \midrule
    \multirow{5}{*}{Hearthstone} & - & 19.70 & 55.96 & 19.76 & 55.14 & 44.49 & 50 min & 47s \\
    ~ & Sequential Integration Fusion & 22.73 & 64.35 & \textbf{14.95} & 63.65 & 60.50 & 60 min & 49s\\
    ~ & Vectorized Decoding Fusion & 22.73 & 64.78 & 16.09 & 65.59 & 60.42 & 320 min & 77s \\
    ~ & Sample Expansion Fusion & 33.33 & 81.10 & 18.82 & 71.71 & 70.29 & 100 min & 48s \\
    ~ & Sketch Filling Fusion & \textbf{34.85} & \textbf{81.89} & 19.71 & \textbf{73.00} &\textbf{71.76} & 107 min & 57s \\
    \bottomrule
  \end{tabular}
  }
  \label{tab:fusion}
\end{table}

\yzz{In addition to evaluating the effects of various fusion strategies on model performance, Table \ref{tab:fusion} also provides insights into the training and inference costs for different fusion strategies on CodeT5. The training time is crucial to understand the computational overhead required by various fusion strategies.
Across all three datasets, SIF emerges as the most time-efficient strategy for training. Compared with SIF, VDF requires a longer duration, which is attributable to the more computational overhead associated with encoding. 
Since both SEF and SFF result in the training samples by a factor of k, they necessitate a training duration that is 2 to 7 times longer than that required for training directly on the original dataset. 
Regarding the inference costs, SEF solely concatenates the most similar code snippet behind the original input, resulting in the shortest input length and the quickest inference time. 
Overall, these fusion strategies exhibit comparable inference times, except for VDF.
Considering both models' performance and costs, SIF is recommended as the fusion strategy in Fusion Phase.}

\begin{center}
\begin{tcolorbox}[colback=gray!10,%gray background
                  colframe=black,% black frame colour
                  width=\linewidth,
                  arc=1mm, auto outer arc,
                  boxrule=1pt,
                 ]
\textbf{Finding 3:} The number of retrieved code snippets should be determined based on the attributes of specific datasets, such as input/output length. 
SIF and SFF can enhance existing pre-trained code models to a greater extent comparing to other fusion strategies. SIF is the most recommended fusion strategy, balancing resource allocation and performance enhancement.
\end{tcolorbox}
\end{center}
\section{Discussion}
\label{chap:chap6}

\begin{table*}
  \caption{Results of LLMs with the RAF on three datasets with different retrieval techniques. The values in parentheses following BLEU and CodeBLEU indicate the ratio of the performance with the RAF compared to the baseline. Variable categories are the same with Table \ref{tab:tab4}.}
  \resizebox{\linewidth}{!}{
  \begin{tabular}{c|c|cc|cc|cc}
    \toprule
    \multirow{2}{*}{\textbf{Model}} & \multirow{2}{*}{\makecell{\textbf{Retrieval}\\ \textbf{Technique}}} & \multicolumn{2}{c|}{\textbf{CONCODE}} & \multicolumn{2}{c|}{\textbf{CoNaLa}} &  \multicolumn{2}{c}{\textbf{HearthStone}} \\
    ~ & ~ &  \textbf{BLEU}  & \textbf{CodeBLEU} & \textbf{BLEU} & \textbf{CodeBLEU} &  \textbf{BLEU} & \textbf{CodeBLEU} \\
    \midrule
    \multirow{6}{*}{ChatGLM} & baseline & 0.06 & 20.21 & 0.31 & 13.43 & 0.03 & 8.05 \\
    ~ & BM25 & 5.76(96.00) & 32.17(1.59)&0.70(2.26) & 21.39(1.59) & \textbf{5.96(198.67)} & \textbf{22.19(2.76)}\\
    ~ & RetroMAE & 2.55(42.50) & 26.86(1.33)& 0.79(2.55) & 21.92(1.63) & 3.44(114.67) & 14.72 (1.83) \\
    ~ & CodeBERT & 5.89(98.17)& 31.02(1.53) & 0.84(2.71)  & 22.36(1.66) & 5.35(178.33) & 19.59(2.43) \\
    ~ & UniXcoder & 5.60(93.33) & 31.30(1.55) & 0.89(2.87) & 20.90(1.56) & 5.31(177.00) & 19.47(2.42) \\
    ~ & \textbf{CoCoSoDa} & \textbf{5.90(98.33)} & \textbf{32.33(1.60)} & \textbf{0.95(3.06)} & \textbf{24.16(1.80)} & 5.11(170.33) & 13.74(1.71)\\
    \midrule
    \multirow{6}{*}{CodeLlama} & baseline & 0.62 & 19.10 & 0.04 & 14.59 & 0.05 & 11.63\\
    ~ & BM25 & 7.45(12.02) & 38.35(2.01) & 0.64(16.00) & 23.88(1.64) & \textbf{7.35(147.00)} & \textbf{47.69(4.10)} \\
    ~ & RetroMAE &  3.72(6.00) & 31.43(1.65) & 0.77(19.25) & 24.02(1.65) & 6.54(130.80) & 42.71(3.67)\\
    ~ & CodeBERT & 7.52(12.13) & 38.70(2.03) & 1.11(27.75) & 25.15(1.72) & 7.03(140.6) & 43.65(3.75) \\
    ~ & UniXcoder & 6.84(11.03) & 38.46(2.01) & 1.11(27.75) & 24.86(1.70) & 6.07(121.40) & 38.94(3.35)\\
    ~ & \textbf{CoCoSoDa} & \textbf{7.54(12.16)} & \textbf{38.86(2.03)} & \textbf{1.25(31.25)} & \textbf{25.93(1.78)} & 6.36(127.20) & 41.43(3.56)\\
    \midrule
    \multirow{6}{*}{DeepSeek-Coder} & baseline & 0.15  & 23.31 & 0.17 & 16.46 & 0.06 & 11.08 \\
    ~ & BM25 & 4.88(32.53) & 37.64(1.61) & 1.09(6.41) & 25.57(1.55) & \textbf{5.44(90.67)} & \textbf{46.21(4.17)}  \\
    ~ & RetroMAE & 2.22(14.80) & 32.82(1.41) & 0.95(5.59) & 25.74(1.56) & 4.41(73.50) & 41.39(3.74) \\
    ~ & CodeBERT & 5.07(33.80) & 37.89(1.63) & 1.12(6.59) & 25.09(1.52) & 4.64(77.33) & 41.26(3.72) \\
    ~ & UniXcoder & 4.68(31.20) & 38.27(1.64) & 1.01(5.94)  & 24.15(1.47) & 4.33(72.17) & 39.45(3.52) \\
    ~ & \textbf{CoCoSoDa} & \textbf{5.45(36.33)} & \textbf{38.75(1.66)}  & \textbf{1.22(7.18)} & \textbf{25.78(1.57)} & 4.94(82.33)  & 41.18(3.72) \\
    \bottomrule
  \end{tabular}
  }
  \label{tab:LLM}
\end{table*}

% \subsection{The impact of Sampling Order for Sequential Integration Fusion}

\subsection{Implications of Findings}
% 检索增强非常有用 如果之前有已经存在的模型可以直接通过增加检索器的方法进一步提升模型的性能
\subsubsection{Implications on the Effectiveness of the Retrieval-Augmented Framework}
The effectiveness and generalization of the retrieval-augmented framework have been proven to improve code generation performance for various existing pre-trained code models.
This implies that it is a workable plan to improve the performance of the model by incorporating retrieval results that are similar to the input, without modifying the model architecture or size.
Diverse models with different architectures can leverage this framework to produce more precise code snippets.
In addition, the retrieval-augmented framework can further enhance the specific model by picking out more powerful retrieval techniques and fusion strategies in the first two phases of this framework. 

% \todo{LLMs}
\yzz{To further investigate the effectiveness of the RAF for large language models (LLMs), we additionally conduct experiments using three popular LLMs: ChatGLM3-6B \cite{chatglm}, CodeLlama-7B \cite{codellama}, and DeepSeek-Coder-6.7B \cite{deepseekcoder}. We also explore the impact of various retrieval techniques for code generation when utilizing these LLMs. The experiment results of the LLMs with RAF on three datasets are shown in Table \ref{tab:LLM}. The retrieved similar code snippets, as references during the code generation process, are integrated into the inputs of the LLMs via Prompt Engineering. The prompts are constructed following \cite{li2024acecoder}. More detailed information about the experiment can be found in our code repository.}

\yzz{As shown in Table \ref{tab:LLM}, all LLMs improve their performance across all three datasets with the RAF. On CONCODE, the BLEU metric ratio, resulting from prompting ChatGLM with the similar codes retrieved by CoCoSoDa, compared to using ChatGLM directly, stands at 98.33. On Hearthstone, the BLEU metric ratio, when using BM25 compared to the original ChatGLM, even escalates to 198.67. These substantial improvements demonstrate that the RAF can effectively enhance the performance of the LLMs in generating target code during the inference phase by providing similar code. This suggests that integrating similar code snippets can greatly aid in the code generation process, leading to more accurate and efficient generation results.}

\begin{table*}
\centering
  \caption{The training costs per epoch and average retrieval costs per 50 instances of different retrieval techniques on different datasets. The retrieved codebases refer to the training set of the corresponding dataset. "-" represents that the retrieval technique does not need to be trained.}
  \resizebox{0.85\linewidth}{!}{
  \begin{tabular}{ccccc}
    \toprule
    \textbf{Dataset} & \textbf{Retrieval Techniques} & \textbf{Training Cost} & \textbf{Retrieval Cost} & \textbf{Total Costs}\\
    \midrule
    \multirow{5}{*}{CONCODE} & BM25 & - & 199.25s & 7970.00s \\
     & RetroMAE & - & 2.02s & 81.00s \\
    % -&-&-&120s生成embedding 推理9 min 45s 585s 5448.55+717.5
    & CodeBERT & 1089.71s & 14.35s & 6166.05s \\
    & UniXcoder & 1680.13s  & 12.50s & 8900.13s \\
    & CoCoSoDa & 2001.02s  & 1.23s & 10051.02s \\
    \midrule
    \multirow{5}{*}{CoNaLa} & BM25 & - & 1.09s & 109.00s \\
     & RetroMAE & - & 4.00s & 40.00s \\
     % 训练6s 生成embedding(no-repeat) 2min4s 
     % 11s生成train的embedding 18s检索结束
    & CodeBERT & 41.13s & 2.90s  & 234.65s \\
    & UniXcoder & 42.00s & 7.67s & 809.00s \\
    & CoCoSoDa & 47.00s & 1.03s & 150.00s\\
    \midrule
    \multirow{5}{*}{HearthStone} & BM25 & - & 1.05s & 1.39s \\
     & RetroMAE & - & 1.52s & 2.00s\\
     % 2min42s 5个epoch 
    & CodeBERT & 32.40s & 5.97s & 169.88s \\
    & UniXcoder & 20.38s  & 15.15s & 115.53s \\
    & CoCoSoDa & 49.47s & 3.79s & 249.26s \\
    \bottomrule
  \end{tabular}
  }
  \label{tab:retrieve_cost}
\end{table*}

\begin{table}
  \caption{Results of the retrieval-augmented CodeT5 for three datasets with different sample ordering. The retrieval technique is BM25, and the fusion strategy is Sequential Integration Fusion. Under each metric, the best performance is highlighted in \textbf{bold}.}
  \resizebox{0.9\linewidth}{!}{
  \begin{tabular}{ccccccc}
    \toprule
    \textbf{Dataset} & \textbf{Sample Ordering Strategy} & \textbf{EM} & \textbf{BLEU} & \textbf{ED} &\textbf{$\text{Sim}_\text{AST}$} & \textbf{CodeBLEU}\\
    \midrule
    \multirow{2}{*}{CONCODE} & \textbf{Ascending Ordering} & \textbf{23.35} & \textbf{40.42} & \textbf{21.91} & \textbf{47.67} & \textbf{46.92} \\
    ~ & Descending Ordering & 22.55 & 36.82 & 24.23 & 47.02 & 45.73 \\
    % ~ & Random Ordering  \\ 
    \midrule
    \multirow{2}{*}{CoNaLa} & \textbf{Ascending Ordering} & \textbf{8.00} & \textbf{13.26} & \textbf{10.46} & \textbf{23.29} & \textbf{21.46}   \\
    ~ & Descending Ordering & 7.20 & 12.18 & 11.32 & 22.84 & 20.43  \\
    % ~ & Random Ordering  \\ 
    \midrule
    \multirow{2}{*}{Hearthstone} & \textbf{Ascending Ordering} & \textbf{22.73} & \textbf{64.35} & \textbf{14.95} & \textbf{63.65} & \textbf{60.50}   \\
    ~ & Descending Ordering & 21.21 & 64.25 & 15.48 & 62.81 & 60.24\\
    % ~ & Random Ordering  \\ 

    \bottomrule
  \end{tabular}
  }
  \label{tab:order}
\end{table}

\subsubsection{Implications on the Utilization of Retrieval-Augmented Framework}
% \todo{the computational overhead, and sample order}
Our experiment results demonstrate that different retrieval techniques in Retrieval Phase influence the performance of retrieval-augmented models for code generation.
\yzz{As shown in Table \ref{tab:tab4} and Table \ref{tab:LLM}, among five retrieval techniques, BM25 and CoCoSoDa perform best with both LLMs and pre-trained models.}
\yzz{To facilitate a clear comparison of the costs associated with various retrieval techniques, we divide the retrieval process costs into two parts: training cost and retrieval cost. The experimental results are presented in Table \ref{tab:retrieve_cost}. For text retrieval algorithms, there is no training cost as no training is required. The total cost is calculated by summing the training cost (if applicable) and the retrieval cost. BM25 is executed on a CPU, while all other deep learning-based retrieval technologies are run on a single A100 GPU to calculate the costs. The total costs indicate that in our experiments, BM25 incurs a smaller cost than the deep learning-based code search models while achieving superior results. However, the advantage of BM25 diminishes as the search volume and input length increases, as evidenced in CONCODE. According to Table \ref{tab:dataset}, the input length of CONCODE is the longest, and the size of the CONCODE training set is the largest among the three datasets. Since BM25 retrieves code by calculating the similarity between natural languages at the token level, its cost on CONCODE is higher than two other datasets. The total costs of BM25 are comparable to deep learning-based retrieval techniques (including training and retrieval). Retrieving similar code snippets for just one instance takes about 4 seconds (199.25s/50), which is intolerable in a real-world code generation scenario. Therefore, BM25 is recommended for datasets with short inputs or small sizes. However, for datasets with longer input/output lengths, more efficient code search models could be crucial.
% should be explored in greater depth. 
Moreover, the experiment results of the LLMs highlight the potential of CoCoSoDa within the RAF, suggesting that future work should delve deeper into exploring more code search models.}

% Sample Expansion Fusion (SEF)
Among various fusion strategies, the Sequential Integration Fusion (SIF) strategy proves to be a simple, direct, and effective fusion strategy.
For the dataset with a clear structure, Sketch Filling Fusion (SFF) has the capacity to improve the model performance, as an optimal fusion strategy. 
% 对于生成器模型的讨论 CodeGen更加敏感 因此选择CodeT5这样更加稳健的模型可能会更好
\yzz{However, this optimal performance comes at a cost: the fusion process of SFF necessitates training the model for k times the sample size. Our experiment results show that their training time is 2-7 times longer than fine-tuning directly on the original dataset. Therefore, when considering both the trade-off between the performance of the fusion strategies and the training costs, SIF is the most recommended.}

\begin{figure}[h]
  \centering
  \includegraphics[width=\linewidth]{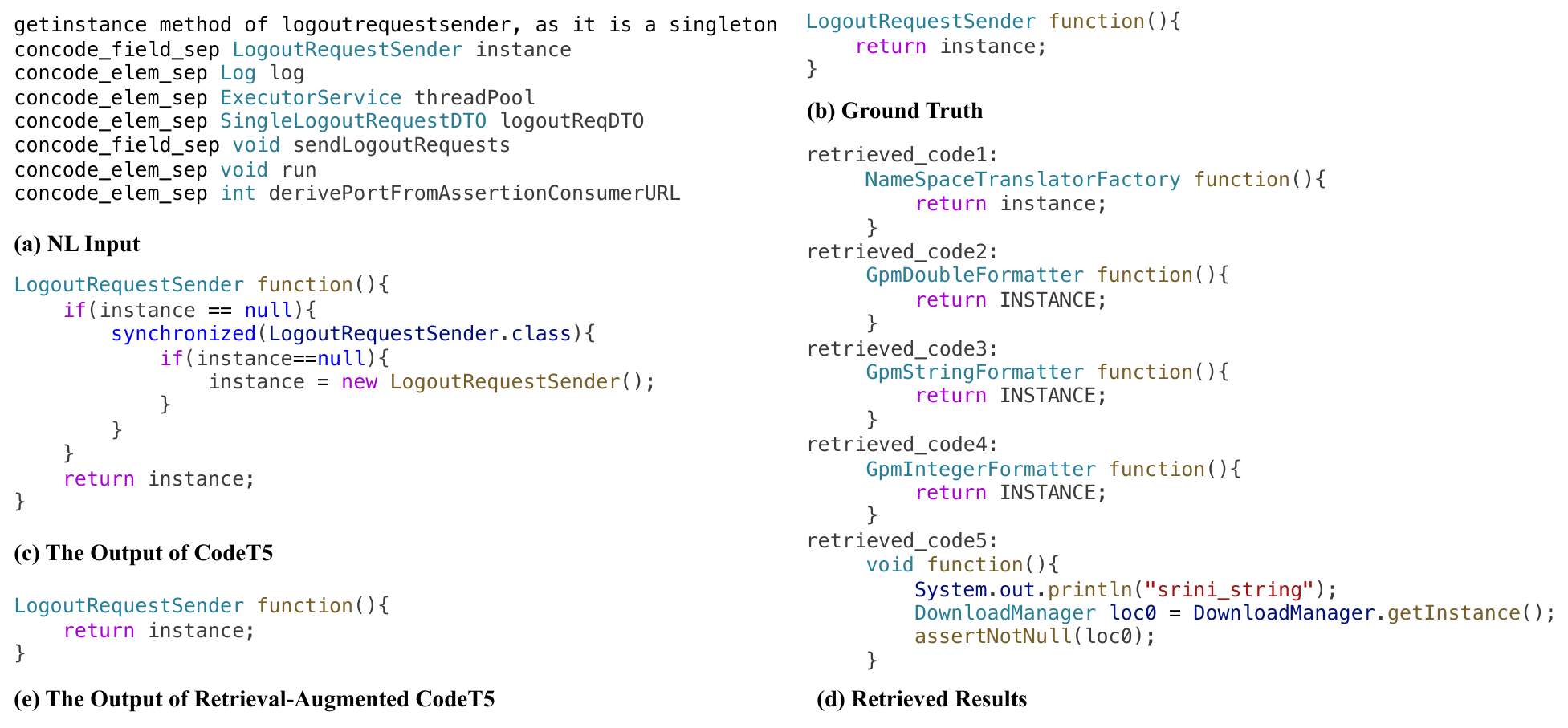}
  \caption{Case study on CONCODE with retrieval-augmented framework, where the retrieval technique is BM25, fusion strategy is Sequential Integration Fusion, and the pre-trained code model is CodeT5.}
  \Description{Case Study on CONCODE with CodeT5}
  \label{fig:fig6-case}
\end{figure}

\begin{figure}[h]
  \centering
  \includegraphics[width=\linewidth]{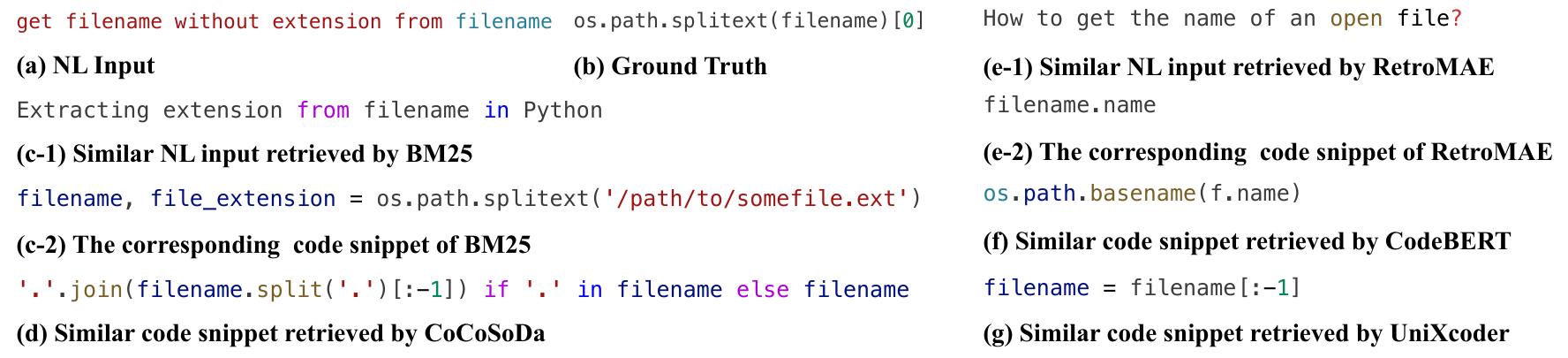}
  \caption{The most similar code snippet retrieved by different retrieval techniques on CoNaLa.}
  \Description{Case Study on CoNaLa with Different Retrieval Techniques.}
  \label{fig:fig6-case-2}
\end{figure}

\yzz{There are two considerations when using SIF: the number and the order of concatenated retrieved code.}
Determining an appropriate number of concatenated retrieved code involves the capacity of the models and the attributes of specific datasets.
% a trade-off between resource allocation and performance enhancement.
It is recommended to opt for a median number based on the length of data in different datasets. 
\yzz{We also conduct experiments with the retrieved code snippets input in ascending ordering (sorted from high to low based on similarity) and in descending ordering (sorted from low to high based on similarity). As indicated in Table \ref{tab:order}, the results demonstrate that the ascending ordering outperforms the descending ordering.}

% 检索增强模型的使用方法，全环节来看，选定一个性能强劲的生成器，其他环节分别讲述  将前面两个问题中的关键点再概括一遍

\subsection{Case Study}
In this section, we provide two case studies to qualitatively compare the base model with the retrieval-augmented model and the retrieved results from different retrieval techniques.
\subsubsection{The effectiveness of the RAF for code generation.}
Figure \ref{fig:fig6-case} shows a case including the original data, the output of original CodeT5, the retrieved results from BM25, and the output of retrieval-augmented CodeT5 on CONCODE.
The requirement is "getinstance methods of logoutrequestsender, as it is a singleton", and the other environment information is shown in Figure \ref{fig:fig6-case} (a).
In Figure \ref{fig:fig6-case} (c), we can observe that the code generated by the original CodeT5 is more complex but does not adhere to the intended specifications.
Indeed, the retrieved results (d) obtained from BM25 exhibit similarity to the ground truth (b). This similarity serves as a valuable reference point for CodeT5, aiding in understanding the genuine intent and determining the appropriate results to be returned.
In this way, retrieval-augmented CodeT5 can generate the same code (e) as ground truth (b).

\subsubsection{Analysis on the retrieved results by different retrieval techniques.}
\yzz{As shown in Figure \ref{fig:fig6-case-2}, different retrieval techniques in the RAF identify various similar code snippets for the subsequent fusion and generation phases. Text retrieval algorithms, such as BM25 and RetroMAE, compute the similarity between the natural language input and natural language description in the retrieved codebase. The corresponding code snippets are then returned based on the retrieved natural language description. Due to the underlying text matching principle, the retrieved natural language description shares many keywords with the input, such as "extension" and "filename" (as shown in Figure \ref{fig:fig6-case-2} (a) and (c-1)). The similar natural language description facilitates the return of similar code snippets related to the ground truth. RetroMAE retrieves natural language descriptions by calculating their semantic similarity. As shown in Figure \ref{fig:fig6-case-2} (e-1), the retrieved natural language description conveys the semantics of obtaining the file name but ignores the condition \textit{``without extension''}, leading to poor retrieval performance. The other three code search models retrieve similar code snippets by directly calculating the similarity between natural language and code snippets. In this case, CoCoSoDa (Figure \ref{fig:fig6-case-2} (d)) and CodeBERT (Figure \ref{fig:fig6-case-2} (f)) retrieve code that meets the requirements of the descriptions, while UniXcoder (Figure \ref{fig:fig6-case-2} (g) does not fully understand the descriptions and returns an intermediate result close to the target answer of CoCoSoDa.}

\subsection{Future Work.}
\yzz{Based on our findings and implications, in this section, we present two possible future works for retrieval-augmented code generation.}

\subsubsection{Active Retrieval.}
\yzz{Our experimental results in Section \ref{chap:5.2} show that not all retrieved code snippets are beneficial for the final generated output. If the retrieved code is irrelevant to the input natural language description, it may confuse code generation models as noise, reducing the generation performance. Additionally, as shown in Table \ref{tab:retrieve_cost}, retrieval incurs extra costs compared to using the code generation model directly. Thus, determining whether retrieval should be performed to improve model performance requires further investigation, which is called active retrieval. Considering that active retrieval improves efficiency and usability in natural language processing tasks \cite{DBLP:conf/emnlp/JiangXGSLDYCN23,DBLP:journals/corr/abs-2406-12534,self-rag} and code-related tasks \cite{DBLP:journals/corr/abs-2403-10059}, we believe it can also optimize and inform the future design of retrieval-augmented framework for code generation tasks.}

\subsubsection{Retrieval Database Construction.}
\yzz{In our experiments and previous works, the retrieval-augmented framework for code generation tasks regards the training set as the retrieval database. However, with the development of general models, evaluation datasets often exclude training sets (e.g., HumanEval \cite{codex} and MBPP \cite{MBPP}). Therefore, constructing a comprehensive, diverse, and informative retrieval database is essential for applying the retrieval-augmented framework to various code generation scenarios. Moreover, based on the database, exploring the potential of fine-tuning deep learning-based models on code generation datasets is a worthwhile endeavor.}

\subsection{Threats to Validity}
% 1. 模型结果的泛化性，尝试的模型不够多 且因为要进行大量实验，因此我们仅选择了较为小规模的预训练模型 但是最近有工作在讨论随着模型规模的增大会出现涌现能力，同时可以使用上下文学习的方法来激活模型的潜力。
\phead{The Generalization of Model Results.}
Code generation stands as one of the most pivotal tasks in code intelligence, where code generation models persistently undergo innovation and evolution.
Given the absence of empirical analysis for retrieval-augmented code generation models, it becomes crucial to conscientiously design a considerable number of experiments to address this research gap.
We select three popular pre-trained models with different architectures for a constrained experimental exploration.
We have made diligent efforts to summarize the experimental results, leading to the discovery of several compelling findings.
Nevertheless, there remains uncertainty regarding whether these findings remain applicable to larger models or models with differing architectures.

% 2. 结果可能的不可复现性质，我们自行完成了代码的检索以及后续增强模型的训练与测试，然而不论是检索器还是后续的生成器，基于深度学习的模型可能会受到不同设备和参数设置的相关影响。直接使用我们的检索结果并复用之前工作的代码和参数设置可以一定程度上保证代码生成的质量。
\phead{The Replication of Our Experiments.}
In this paper, we perform diverse retrieval techniques along with training and testing of the augmented models.
Nevertheless, it is essential to note that both the code search models and the pre-trained code models based on deep learning could be influenced by various factors, including different devices and parameter settings.
To address this issue, we have made our retrieval-augmented datasets\footnote{\url{https://drive.google.com/drive/folders/1G\_ssf9gCX38Yb7FAjsIlRxiYPO44omvP?usp=drive\_link}} and code repository\footnote{\url{https://github.com/watreyoung/RACG}} publicly available.
By utilizing our retrieval results directly, along with the code and parameter settings from previous work, the consistency of the code generation tasks with our experiment results will be ensured to some extent.

% 3. 与现实场景之间的差异，数据集的组织形式和真实开发环境仍然存在一定的差距，但是这样的情况普遍存在于现在以模型为核心的方法中
\phead{Limited Dataset.}
The experiment results are based on three datasets for code generation task.
While we have selected the most widely used and representative datasets for our experiment, there remains a distinct gap between the data within these datasets and the real, specific development environment context.
For example, CONCODE is the most used benchmark for code generation task, but the pre-processing makes human hard to understand the code intuitively.
In short, it is difficult for human developers to write code the same as the ground truth according to the natural language description on CONCODE.
This issue persists across numerous tasks within code intelligence, indicating a potential future research direction to construct a dataset that better mirrors real-world development scenarios.

\section{Conclusion}
\label{chap:chap7}
In this paper, we experimentally investigate the effectiveness and generalization of the retrieval-augmented framework for code generation on three different datasets.
Our study shows that retrieval-augmented framework can indeed improve the code generation performance of existing code pre-trained models, such as CodeGen, UniXcoder, and CodeT5.
% for code generation.
Besides, we explore the impact of different retrieval techniques and different fusion strategies on retrieval-augmented framework.
Plenty of experimental results are listed and discussed in section \ref{chap:chap5}.
We summarize our findings and provide some implications for the utilization of retrieval-augmented framework for code generation.
These insights may assist researchers in leveraging the retrieval-augmented framework to enhance their own models for code generation.

\normalem
\bibliographystyle{ACM-Reference-Format}
\bibliography{main}

\end{document}